\documentclass[pre,aps,preprint,showpacs,preprintnumbers,amsmath,amssymb,secnumroman,eqsecnum,eqappnum]{revtex4}

\usepackage{graphicx}
\usepackage{dcolumn}
\usepackage{bm}

\newcommand{\beq}{\begin{equation}}
\newcommand{\eeq}{\end{equation}}
\newcommand{\beqa}{\begin{eqnarray}}
\newcommand{\eeqa}{\end{eqnarray}}
\newcommand{\vc}[1]{\mbox{\boldmath $#1$}}

\newcommand{\vol}[1]{{\bf #1}}

\newcommand{\du}[1]{{\bf\sf #1}}

\begin{document}


\title{Effect of fluid inertia on swimming of a sphere in a viscous incompressible fluid}

\author{B. U. Felderhof}

 \email{ufelder@physik.rwth-aachen.de}
\affiliation{Institut f\"ur Theorie der Statistischen Physik\\ RWTH Aachen University\\
Templergraben 55\\52056 Aachen\\ Germany\\
}%

\author{R. B. Jones}

 \email{r.b.jones@qmul.ac.uk}
\affiliation{Queen Mary University of London, The School of
Physics and Astronomy, Mile End Road, London E1 4NS, UK\\}%

\date{\today}

\begin{abstract}
Swimming of a sphere in a viscous incompressible fluid is studied on the basis of the Navier-Stokes equations for wave-type distortions of the spherical shape. At sizable values of the dimensionless scale number the mean swimming velocity is the result of a delicate balance between the net time-averaged flow generated directly by the surface distortions and the flow generated by the mean Reynolds force density. Depending on the stroke, this can lead to a surprising dependence of the mean swimming velocity on the kinematic viscosity of the fluid. The net flow pattern is calculated as a function of kinematic viscosity for axisymmetric strokes of the swimming sphere. The calculation covers the full range of scale number, from the friction-dominated Stokes regime in the limit of vanishing scale number to the inertia-dominated regime at large scale number. The model therefore provides paradigmatic insight into the fluid dynamics of swimming or flying of a wide range of organisms.
\end{abstract}

\maketitle
\section{\label{I}Introduction}

The theory of swimming of nearly spherical microorganisms was developed first by Lighthill \cite{1} and Blake \cite{2}. The Reynolds number for a microorganism is small and the theory can be based on the time-independent Stokes equations \cite{3},\cite{4}. Work in this area was reviewed by Lauga and Powers \cite{5}.

In the swimming of larger bodies inertia of the fluid can no longer be neglected. The Reynolds stress causes a reactive flow which modifies the Stokes flow. In earlier work we discussed the effect of fluid inertia on swimming on the basis of the complete set of Navier-Stokes equations \cite{6}. As the simplest geometry we studied the swimming of a sphere with no-slip boundary condition due to time-harmonic surface distortions \cite{7}. The swimming velocity was evaluated to second order in the amplitude of distortions as an average over a period of time. In terms of the dimensionless ratio $\varepsilon=\xi/a$, where $\xi$ is the amplitude of stroke and $a$ is the ratio of the undistorted sphere, the mean swimming velocity and the mean rate of dissipation were calculated to order $\varepsilon^2$.

The effect of fluid inertia may be characterized by the dimensionless ratio $s^2=a^2\omega\rho/(2\eta)$, where $\omega$ is the frequency of the stroke, $\rho$ is the mass density of the fluid, and $\eta$ is the shear viscosity. For fixed $a$ and $\omega$ one may consider the full range of kinematic viscosity $\nu=\eta/\rho$. The Stokes limit corresponds to $s=0$. In this limit the Reynolds stress vanishes. For larger values of the scale number $s$ the Reynolds stress becomes important.

In earlier work we calculated the mean swimming velocity and the mean rate of dissipation of a swimming sphere in the full range of scale number \cite{8}.
For finite values of $s$ the theory employs mode functions which become singular in the limit $s=0$. This causes mathematical difficulties which obscure the relation to the Stokes limit. In our earlier work \cite{8} we used a modified set of mode functions for small $s$ to facilitate the study of low frequency behavior. We show here that the theory can be improved by a matrix representation based on the surface modes of the Stokes limit theory.

The new representation allows us to study the mean swimming velocity for a choice of stroke which is the same for all values of the scale number $s$. As a consequence the dependence of the mean swimming velocity on scale number is due entirely to the variation of mass density and viscosity of the fluid. It turns out that the swimming velocity has a surprising dependence on $s$. For certain strokes the swimming velocity changes sign as $s$ increases.

The calculations show that for larger values of $s$ there is a delicate balance between the direct effect of surface distortion and the indirect effect of the Reynolds stress. The two contributions to the mean swimming velocity, generated directly and indirectly, nearly cancel for large $s$, whereas for small $s$ the indirect contribution vanishes.

Our earlier work \cite{8} was limited to a calculation of the mean swimming velocity as a function of the stroke and of fluid properties. In the present article we study in addition the mean flow pattern of a swimmer, the mean being defined as the time-average over a period. It is interesting to see for fixed chosen stroke how the flow pattern changes as the kinematic viscosity decreases from a high value in the friction-dominated Stokes regime to a low value in the inertia-dominated regime.

We view the effect of swimming as force-free convection of a body in self-generated fluid flow. It is of interest to study the net time-averaged flow as a superposition of the direct and indirect contributions. Both contributions depend on the scale number $s$, and again for large $s$ the two contributions nearly cancel, whereas for small $s$ the indirect contribution tends to zero. We show in examples that for fixed stroke the net flow pattern varies significantly as a function of scale number. The calculations are exact to order $\varepsilon^2$. We emphasize that the calculation covers the full range of scale number.

The calculation is conceptually important. Although the sphere geometry is rather special, it has the advantage that the calculation on the basis of the Navier-Stokes equations can be performed in full analytic detail. The calculation turns out to be surprisingly complex. It exhibits the subtle interplay of friction and inertia in the fluid dynamics of swimming, and can serve as a guide in the analysis of more realistic geometries, such as a distorting spheroid.

The effect of Reynolds stress on the translational velocity of a $B_1B_2$-active particle, a so-called squirmer, was studied by Wang and Ardekani \cite{9}, by Khair and Chisholm \cite{10}, and by Chisholm et al. \cite{11}. Spelman and Lauga \cite{12} studied the translational velocity of a squirmer in the inertia-dominated limit to order $\varepsilon^2$ by the method of matched asymptotic expansion. Wang and Ardekani \cite{13} studied the effect of fluid inertia on swimming of small organisms via an approximate equation of motion.

\section{\label{II}Swimming of a sphere}

We consider a flexible sphere of radius $a$ immersed in a viscous
incompressible fluid of shear viscosity $\eta$ and mass density $\rho$.
The fluid is set in motion by time-dependent distortions of the
sphere. We shall study axisymmetric periodic distortions which lead to a translational swimming
motion of the sphere in the $z$ direction in a Cartesian system of coordinates. The analysis is based on a perturbation expansion of the Navier-Stokes equations in powers of the amplitude of distortions \cite{6}. In the following the mean swimming velocity, the mean rate of dissipation, and the mean flow pattern are evaluated to second order in the amplitude. The no-slip boundary condition is applied on the surface of the distorted sphere. For a complete discussion of the equations we refer to earlier work \cite{6},\cite{8}.

The surface displacement is written as
\begin{equation}
\label{2.1}
\vc{\xi}(\theta,t)=\mathrm{Re}[\vc{\xi}_\omega(\theta)e^{-i\omega t}],
\end{equation}
with polar angle $\theta$ and complex amplitude $\vc{\xi}_\omega(\theta)$. The corresponding first order flow velocity and pressure are given by
\begin{equation}
\label{2.2}
\vc{v}^{(1)}(\vc{r},t)=\mathrm{Re}[\vc{v}_\omega(\vc{r})e^{-i\omega t}],\qquad p^{(1)}(\vc{r},t)=\mathrm{Re}[p_\omega(\vc{r})e^{-i\omega t}],
\end{equation}
with amplitude functions which satisfy the linearized Navier-Stokes equations
\begin{equation}
\label{2.3}\eta[\nabla^2\vc{v}_\omega-\alpha^2\vc{v}_\omega]-\nabla p_\omega=0,\qquad\nabla\cdot\vc{v}_\omega=0,
\end{equation}
with the variable
\begin{equation}
\label{2.4}\alpha=(-i\omega\rho/\eta)^{1/2}=(1-i)(\omega\rho/2\eta)^{1/2}.
\end{equation}
The solution of Eq. (2.3) can be expressed as a linear superposition of modes \cite{14}
 \begin{eqnarray}
\label{2.5}\vc{v}_l(\vc{r},\alpha)&=&\frac{2}{\pi}\;e^{\alpha a}[(l+1)k_{l-1}(\alpha r)\vc{A}_l(\hat{\vc{r}})+lk_{l+1}(\alpha r)\vc{B}_l(\hat{\vc{r}})],\nonumber\\
\vc{u}_l(\vc{r})&=&-\bigg(\frac{a}{r}\bigg)^{l+2}\vc{B}_l(\hat{\vc{r}}),\qquad p_l(\vc{r},\alpha)=\eta\alpha^2a\bigg(\frac{a}{r}\bigg)^{l+1}P_l(\cos\theta),
\end{eqnarray}
with modified spherical Bessel functions \cite{15} $k_l(z)$ and vector spherical harmonics $\{\vc{A}_l,\vc{B}_l\}$ defined by \cite{16}
 \begin{eqnarray}
\label{2.6}\vc{A}_l&=&\hat{\vc{A}}_{l0}=lP_l(\cos\theta)\vc{e}_r-P^1_l(\cos\theta)\vc{e}_\theta,\nonumber\\
\vc{B}_l&=&\hat{\vc{B}}_{l0}=-(l+1)P_l(\cos\theta)\vc{e}_r-P^1_l(\cos\theta)\vc{e}_\theta,
\end{eqnarray}
with Legendre polynomials $P_l$ and associated Legendre functions $P^1_l$ in the notation of Edmonds \cite{17}.
The corresponding surface displacement function $\vc{\xi}_\omega(\hat{\vc{r}})$  is expanded as
\begin{equation}
\label{2.7}\vc{\xi}_\omega(\hat{\vc{r}})=-ia\sum^\infty_{l=1}[\kappa_l\vc{v}_l(\vc{s},\alpha)+\mu_l\vc{u}_l(\vc{s})],
\end{equation}
with $\vc{s}=a\hat{\vc{r}}$ and complex coefficients $\{\kappa_l,\mu_l\}$. The first order fluid velocity at the surface is given by $\vc{v}^{(1)}(\vc{s},t)=\partial\vc{\xi}(\hat{\vc{r}},t)/\partial t$.

The mean second order flow velocity $\overline{\vc{v}^{(2)}}$ and pressure $\overline{p^{(2)}}$ satisfy the inhomogeneous Stokes equations \cite{6}
\begin{equation}
\label{2.8}
\eta\nabla^2\overline{\vc{v}^{(2)}}-\nabla\overline{p^{(2)}}=\frac{1}{2}\rho\;\mathrm{Re}\;[\vc{v}^*_\omega\cdot\nabla\vc{v}_\omega],\qquad\nabla\cdot\overline{\vc{v}^{(2)}}=0,
\end{equation}
with boundary condition
\begin{equation}
\label{2.9}
\overline{\vc{v}^{(2)}}\big|_{r=a}=\overline{\vc{u}}_S(\theta)=-\frac{1}{2}\mathrm{Re}\;[\vc{\xi}^*_\omega\cdot\nabla\vc{v}_\omega]\bigg|_{r=a}.
\end{equation}
The mean is defined as the time-average over a period $T=2\pi/\omega$.
The right hand side in Eq. (2.8) represents the mean Reynolds force density $\overline{\vc{f}^{(2)}_R}=-\rho\overline{\vc{v}^{(1)}\cdot\nabla\vc{v}^{(1)}}$. The volume part of the second order flow $\overline{\vc{v}^{(2)}_{V}},\overline{p^{(2)}_{V}}$ satisfies Eq. (2.8) with the no-slip boundary condition $\overline{\vc{v}^{(2)}}_V\big|_{r=a}=0$. The surface part $\overline{\vc{v}^{(2)}_{S}},\overline{p^{(2)}_{S}}$ satisfies Eq. (2.8) with right hand side equal to zero and with boundary condition Eq. (2.9).

We define the complex multipole moment vector $\vc{\psi}$ as the one-dimensional array of coefficients in Eq. (2.7),
\begin{equation}
\label{2.10}\vc{\psi}=(\kappa_1,\mu_1,\kappa_2,\mu_2,....).
\end{equation}
The absence of uniform displacement implies the constraint $\kappa_1=0$. We indicate arrays satisfying this constraint by a hat, as $\hat{\vc{\psi}}$. The mean swimming velocity
$\overline{U}_2$ and the mean rate of dissipation $\overline{\mathcal{D}}_2$ are bilinear in the vector $\hat{\vc{\psi}}$ and can be expressed as \cite{6}
\begin{equation}
\label{2.11}\overline{U_2}=\frac{1}{2}\omega
a(\hat{\vc{\psi}}|\hat{\du{B}}|\hat{\vc{\psi}}),\qquad\overline{\mathcal{D}_2}=8\pi\eta\omega^2a^3
(\hat{\vc{\psi}}|\hat{\du{A}}|\hat{\vc{\psi}}).
\end{equation}
With truncation at maximum $l$-value $L$ the
truncated matrices $\hat{\du{A}}_{1L}$ and $\hat{\du{B}}_{1L}$ are $2L-1$-dimensional. The truncated matrices correspond to swimmers obeying the constraint that all multipole coefficients for $l>L$ vanish.

The matrices $\hat{\du{A}}$ and $\hat{\du{B}}$ are calculated from integrals with integrands which are bilinear in the mode functions defined in Eq. (2.5). The matrix $\hat{\du{A}}$ is diagonal in $l,l'$, and the matrix $\hat{\du{B}}$ is tridiagonal in $l,l'$. The matrices are frequency-dependent via the variable $\alpha a$. We write
\begin{equation}
\label{2.12}\alpha a=(1-i)s,\qquad s=a\sqrt{\frac{\omega\rho}{2\eta}}
\end{equation}
with dimensionless scale number $s$. In Appendix B of Ref. 8 we provided explicit expressions for the matrix elements of $\hat{\du{A}}(s)$ and $\hat{\du{B}}(s)$ up to $L=3$.

\section{\label{III}Stokes representation}

The matrices $\hat{\du{A}}(s)$ and $\hat{\du{B}}(s)$ are singular at $s=0$ which causes difficulties in numerical calculations and in the discussion of the relation to swimming in the Stokes limit. In our earlier work \cite{8} we therefore used for small $s$ a different set of mode solutions of Eq. (2.3). However, alternatively we can choose a more convenient matrix representation by expanding the surface displacement $\vc{\xi}_\omega(\hat{\vc{r}})$ in terms of a different set of vector functions defined on the surface of the sphere $r=a$. It is of particular interest to use the set of functions found as limiting values on the sphere surface of the modes defined in the Stokes limit \cite{18}. The mode functions $\vc{u}_l(\vc{r})$ in the Stokes limit are the same as in Eq. (2.5), but the functions $\vc{v}_l(\vc{r},\alpha)$ are changed to
\begin{eqnarray}
\label{3.1}\vc{v}^0_l(\vc{r})&=&\bigg(\frac{a}{r}\bigg)^l\bigg[(l+1)P_l(\cos\theta)\vc{e}_r+\frac{l-2}{l}P^1_l(\cos\theta)\vc{e}_\theta\bigg]\nonumber\\
&=&\bigg(\frac{a}{r}\bigg)^l\bigg[\frac{2l+2}{l(2l+1)}\vc{A}_l-\frac{2l-1}{2l+1}\vc{B}_l\bigg].
\end{eqnarray}
We denote the corresponding set of superposition coefficients as $\hat{\vc{\psi}}^I=(\mu_1^I,\kappa_2^I,\mu_2^I,...)$ and the corresponding Stokes representation of the matrices as $\hat{\du{A}}^I(s)$ and $\hat{\du{B}}^I(s)$. The Stokes limit is denoted as $\hat{\du{A}}^0=\hat{\du{A}}^I(0)$ and $\hat{\du{B}}^0=\hat{\du{B}}^I(0)$.

In our earlier work \cite{8} we gave the relation between the two sets of mode coefficients  $(\mu_1,\kappa_2,\mu_2)$ and $(\mu_1^I,\kappa_2^I,\mu_2^I)$, found from a comparison of the two expansions of the surface displacement on the sphere. More generally we find by the same method the linear relation
\begin{equation}
\label{3.2}\hat{\vc{\psi}}=\du{T}\cdot\hat{\vc{\psi}}^I,
\end{equation}
with a transformation matrix $\du{T}$. The matrix $\du{T}$ is block-diagonal, as given by a factor $\delta_{ll'}$, with a 2-dimensional $\du{T}_l$ at order $l$ given by the relations
\begin{eqnarray}
\label{3.3}\kappa_l&=&\frac{\pi}{l(2l+1)e^zk_{l-1}(z)}\;\kappa^I_l,\nonumber\\
\mu_l&=&\frac{1}{2l+1}\bigg[2l-1+2\frac{k_{l+1}(z)}{k_{l-1}(z)}\bigg]\kappa^I_l+\mu^I_l,\qquad z=(1-i)s.
\end{eqnarray}
At $l=1$ we have simply $\mu_1=\mu^I_1$. The relation between the two sets of matrices is
\begin{equation}
\label{3.4}\hat{\du{A}}^I=\du{T}^\dagger\cdot\hat{\du{A}}\cdot\du{T},\qquad\hat{\du{B}}^I=\du{T}^\dagger\cdot\hat{\du{B}}\cdot\du{T},
\end{equation}
where $\du{T}^\dagger$ is the Hermitian conjugate of $\du{T}$. The mean swimming velocity
$\overline{U}_2$ and the mean rate of dissipation $\overline{\mathcal{D}}_2$ can be expressed alternatively as
\begin{equation}
\label{3.5}\overline{U_2}=\frac{1}{2}\omega
a(\hat{\vc{\psi}}^I|\hat{\du{B}}^I|\hat{\vc{\psi}}^I),\qquad\overline{\mathcal{D}_2}=8\pi\eta\omega^2a^3
(\hat{\vc{\psi}}^I|\hat{\du{A}}^I|\hat{\vc{\psi}}^I).
\end{equation}

The explicit expression for the matrix $\hat{\du{A}}^I$ up to order $L=3$ reads
\begin{equation}
\label{3.6}\hat{\du{A}}^I_{13}=\left(\begin{array}{ccccc}3&0&0&0&0
\\0&A^I_{22}&\frac{18}{5}&0&0
\\0&\frac{18}{5}&6&0&0
\\0&0&0&A^I_{44}&\frac{50}{7}
\\0&0&0&\frac{50}{7}&10\end{array}\right),
\end{equation}
with elements
\begin{eqnarray}
\label{3.7}A^I_{22}&=&\frac{3}{10}\;\frac{9+18s+18s^2+2s^3}{1+2s+2s^2},\nonumber\\
A^I_{44}&=&\frac{2}{21}\;\frac{531+1062s+1062s^2+708s^3+244s^4+8s^5}{9+18s+18s^2+12s^3+4s^4}.
\end{eqnarray}
The matrix $\hat{\du{A}}^I_{13}$ is simpler than $\hat{\du{A}}_{13}$. It differs only in the $22$- and $44$-elements from the matrix $\hat{\du{A}}^0_{13}$, which can be read off from Eq. (7.17) in Ref. 18.

The matrix $\hat{\du{B}}^I$ is a sum of two terms $\hat{\du{B}}^I=\du{B}^I_S+\du{B}^I_B$, corresponding to contributions from surface displacements directly and from a bulk term originating in the Reynolds stress tensor. The nonvanishing elements $B^I_{S\alpha\beta}$ are given by
 \begin{eqnarray}
\label{3.8}B^I_{S12}&=&B^{I*}_{S21}=\frac{3+(3-3i)s+4is^2}{5i+(5+5i)s},\nonumber\\
B^I_{S13}&=&B^{I*}_{S31}=-3i,\nonumber\\
B^I_{S24}&=&B^{I*}_{S42}=\frac{6}{35}\;\frac{9+18s+(18-14i)s^2-(6+10i)s^3-(8-8i)s^4}{3i+6is+(2+6i)s^2+(2+2i)s^3},\nonumber\\
B^I_{S25}&=&B^{I*}_{S52}=\frac{6}{35}\;\frac{-15i+(15-15i)s-4s^2}{1+(1+i)s},\nonumber\\
B^I_{S34}&=&B^{I*}_{S43}=\frac{6}{35}\frac{45+(45-45i)s-26is^2+(4+4i)s^3}{3i+(3+3i)s+2s^2},\nonumber\\
B^I_{S35}&=&B^{I*}_{S53}=-6i.
\end{eqnarray}
The matrix $\hat{\du{B}}^I_{S13}$ is simpler than $\hat{\du{B}}_{S13}$.

The nonvanishing elements $B^I_{B\alpha\beta}$ are given by
 \begin{eqnarray}
\label{3.9}B^I_{B12}&=&B^{I*}_{B21}=\frac{s^2}{5}\;\frac{-i-(1+i)s+s^2-(1-i)s^3-2is^4F_-}{i+(1+i)s},\nonumber\\
B^I_{B24}&=&B^{I*}_{B42}=\frac{s^2}{630[3i+6is+(2+6i)s^2+(2+2i)s^3]}\times\nonumber\\
&\big[&-216i-432is-(234+432i)s^2+(198+378i)s^3\nonumber\\&-&12i(7-56i+96F_2-45F_++21F_-)s^4+(6+6i)(-20+33i+90F_+-42iF_-)s^5,\nonumber\\
&+&(-51+124i-384F_2+585F_++255F_-)s^6+(1+i)(55+6i-225iF_++87F_-)s^7\nonumber\\
&+&(16+3i-117iF_++3iF_-)s^8-(1-i)(-3+15iF_++F_-)s^9-6F_+s^{10}\big],\nonumber\\
B^I_{B25}&=&B^{I*}_{B52}=\frac{s^2}{420(1+(1+i)s)}\;\big[18+(18+18i)s+6is^2+(6-6i)s^3-9s^4\nonumber\\&+&11(1+i)s^5+i(1-24F_+)s^6+(1-i)s^7-2F_+s^8\big],\nonumber\\
B^I_{B34}&=&B^{I*}_{B43}=\frac{s^2}{1260[3i+(3+3i)s+2s^2]}\times\nonumber\\
&\big[&-450i-(450+450i)s-222s^2+(78-78i)s^3+81is^4\nonumber\\&-&(83+83i)s^5-(1-168F_-)s^6+(1-i)s^7+2iF_-s^8\big],
\end{eqnarray}
with the abbreviations
\begin{equation}
\label{3.10}F_+=F(s+is),\qquad F_-=F(s-is),\qquad F_2=F(2s),
\end{equation}
where the function $F(z)$ with complex variable $z$ is defined by
\begin{equation}
\label{3.11}F(z)=e^zE_1(z)=\int^\infty_0\frac{e^{-u}}{z+u}\;du.
\end{equation}
As we shall see, the contributions from the matrix $\du{B}^I_B$ are important for large $s$. In the limit $s=0$ the matrix $\du{B}^I_{B13}$ vanishes, and the matrix $\du{B}^I_{S13}$ tends to the matrix $\du{B}^0_{13}$ which can be read off from Eq. (7.11) in Ref. 18.

\section{\label{IV}Simple swimmers}

We study the effect of fluid inertia on swimming performance by calculating the mean swimming velocity $\overline{U_2}$ as a function of scale number $s$ for fixed surface displacement $\vc{\xi}_\omega(\hat{\vc{r}})$. This implies a fixed set of coefficients $\{\kappa_l^I,\mu_l^I\}$. In order to compare different swimmers we define the dimensionless reduced swimming velocity
\begin{equation}
\label{4.1}U_{red}(s)=\frac{(\hat{\psi}^I|\hat{\du{B}}^I(s)|\hat{\psi}^I)}{(\hat{\psi}^I|\hat{\du{A}}^0|\hat{\psi}^I)}.
\end{equation}
The denominator provides a measure of the intensity of surface agitation. For a chosen fixed set of coefficients $\{\kappa_l^I,\mu_l^I\}$ it is independent of scale number $s$.

In the following we consider the simplest swimmers involving only modes of orders $l=1,2,3$. For such swimmers the reduced swimming velocity $U_{red}(s)$ can be evaluated from the explicit expressions for the matrix elements of $\hat{\du{B}}^I_{13}(s)$ given in Sec. III. The matrix $\hat{\du{A}}^0_{13}$ can be read off from Eq. (7.17) in Ref. 18, or from Eq. (3.7) with $s=0$.

The simplest swimmer is the potential one, superposing a dipolar and a quadrupolar flow field, corresponding to moments $\mu_1=\mu^I_1=1$, $\mu_2=\mu^I_2=i/\sqrt{2}$, and all other moments vanishing. For this swimmer the reduced swimming velocity is $U_{red}=1/\sqrt{2}=0.701$, independent of $s$. We have optimized the ratio of the two moments. With three optimized moments $\mu_1=1,\;\mu_2=i\sqrt{11/10},\;\mu_3=-3/5$ the swimming velocity increases to $U_{red}=\sqrt{11/10}=1.049$.

Next we study the so-called $B_1B_2$-swimmer, as defined in terms of the modes introduced by Lighthill \cite{1} and Blake \cite{2}. In terms of the mode coefficients defined above
\begin{equation}
\label{4.2}\mu^I_1=B_1,\qquad\kappa^I_2=-\mu^I_2,\qquad\mu^I_2=\frac{1}{3}iB_2.
\end{equation}
In our scheme the first order swimming velocity $U_1$ vanishes, so that Lighthill's coefficient $A_1$ equals $2\mu^I_1$.
The constraint $\kappa^I_2=-\mu^I_2$ causes the first order $l=2$ component of the flow to be tangential to the sphere. We consider in particular the case $B_2/B_1=5$, the same ratio as for the $B_1B_2$-active particle studied by Ishikawa et al. \cite{19}. In Fig. 1 we plot the reduced swimming velocity $U_{red}(s)$ as a function of $s$. In the Stokes limit $U_{red}(0)=48/43=1.116$. For large scale number $U_{red}(\infty)=72/43=1.674$.

In Fig. 1 we compare with the reduced swimming velocity of the 12-swimmer with optimal velocity in the Stokes limit, as given by the solution of the generalized eigenvalue problem \cite{18} with matrices $\hat{\du{B}}^0_{12}$ and $\hat{\du{A}}^0_{12}$. This corresponds to moments $(\mu^I_1,\kappa^I_2,\mu^I_2)=(1,-4i\sqrt{2}/3,11i/(5\sqrt{2}))$. In the Stokes limit $U_{red}(0)=5/(3\sqrt{2})=1.179$. For large scale number $U_{red}(\infty)=41/(15\sqrt{2})=1.933$.

There are no diagonal elements in Eqs. (3.8) and (3.9). This implies that the stroke must contain at least two different modes of the chosen type. Another simple swimmer has only the coefficients $\mu^I_1$ and $\kappa^I_2$ different from zero. Optimizing again in the Stokes limit we obtain the values $\mu^I_1=1,\kappa^I_2=i\sqrt{10}/3$. In Fig. 2 we plot the reduced swimming velocity $U_{red}(s)$ for this swimmer. In the Stokes limit $U_{red}(0)=\sqrt{2/45}=0.211$. For large scale number $U_{red}(\infty)=-\sqrt{2/45}$. Remarkably, the reduced swimming velocity $U_{red}(s)$ changes sign as a function of $s$. The swimmer is not very efficient.

In Fig. 2 we also plot the reduced swimming velocity $U_{red}(s)$ for a 123-swimmer with so-called combined stroke \cite{6}. The optimized coefficients are
\begin{equation}
\label{4.3}\mu^I_1=1,\qquad\kappa^I_2=\frac{5}{3}\sqrt{\frac{230}{413}}\;i,\qquad\mu^I_2=0,\qquad
\kappa^I_3=-\frac{27}{59},\qquad\mu^I_3=0,
\end{equation}
as calculated from the Stokes generalized eigenvalue problem with constraints $\mu^I_2=0,\;\mu^I_3=0$. Again, in this case the reduced swimming velocity changes sign as a function of $s$. In the Stokes limit $U_{red}(0)=0.249$. For large scale number $U_{red}(\infty)=-0.607$.

The sign change occurs also for a 23-swimmer with moments
\begin{equation}
\label{4.4}\mu^I_1=0,\qquad\kappa^I_2=1,\qquad\mu^I_2=0,\qquad
\kappa^I_3=\frac{9}{2}\sqrt{\frac{7}{295}}\;i,\qquad\mu^I_3=0,
\end{equation}
as shown in Fig. 2. Again we optimized the moments in the Stokes limit. The corresponding reduced velocity is $U_{red}(0)=0.132$. For large $s$ the velocity tends to $U_{red}(\infty)=-0.807$.

Finally we consider the 123-swimmer with optimized moments in the Stokes limit. The moments are
\begin{equation}
\label{4.5}\mu^I_1=1,\qquad\kappa^I_2=-1.553i,\qquad\mu^I_2=1.824i,\qquad
\kappa^I_3=1.373,\qquad\mu^I_3=-1.440,
\end{equation}
corresponding to Stokes reduced velocity $U_{red}(0)=1.514$. For large $s$ the velocity tends to $U_{red}(\infty)=2.175$. In Fig. 1 we show the reduced velocity $U_{red}(s)$ as a function of $s$.

\section{\label{V}Analysis of mean swimming velocity}

In this section we analyze the results shown above in some more detail. The potential swimmer needs no further comment. The reduced velocity of the $B_1B_2$-swimmer can be expressed as
\begin{equation}
\label{5.1}U_{red}(s)=\frac{4\beta}{90+5\beta^2}\frac{12+24s+24s^2+4s^3-s^4+2s^6-is^6(1+s-is)F_++is^6(1+s+is)F_-]}{1+2s+2s^2},
\end{equation}
where $\beta=B_2/B_1$. In particular for $s=0$ and for $s\rightarrow\infty$
\begin{equation}
\label{5.2}U_{red}(0)=\frac{48\beta}{90+5\beta^2},\qquad U_{red}(\infty)=\frac{72\beta}{90+5\beta^2}.
\end{equation}
In the literature a Stokes active particle with $\beta>0$ has been called a puller, and one with $\beta<0$ a pusher \cite{19}. For a Stokes $B_1B_2$-swimmer, as defined here, the names are inappropriate, since one deals with the same swimmer swimming in opposite directions. The reduced velocity of the swimmer is maximal for $\beta=3\sqrt{2}=4.243$  at any $s$. In the Stokes limit $U_{red}(0)=1.131$. The distinction between Stokes active particle and Stokes swimmer was discussed by one of us \cite{20}.

The same type of factorization as shown in Eq. (5.1) occurs for the swimmer of type considered in Eq. (4.4). More generally we put
\begin{equation}
\label{5.3}\mu^I_1=0,\qquad\kappa^I_2=1,\qquad\mu^I_2=0,\qquad
\kappa^I_3=i\gamma,\qquad\mu^I_3=0.
\end{equation}
The reduced swimming velocity takes the form
\begin{equation}
\label{5.4}U_{red}(s)=\frac{216\gamma}{567+1180\gamma^2}\;G(s),
\end{equation}
where the function $G(s)$ takes the limiting values
\begin{equation}
\label{5.5}G(0)=1,\qquad G(\infty)=-\frac{55}{9}.
\end{equation}
The prefactor takes its maximum at the value $\gamma$ given in Eq. (4.4).

In general the reduced swimming velocity can be written as a sum of two terms
\begin{equation}
\label{5.6}U_{red}(s)=U_S(s)+U_B(s),
\end{equation}
corresponding to the decomposition $\hat{\du{B}}^I=\du{B}^I_S+\du{B}^I_B$ mentioned above Eq. (3.8). The bulk term $U_B(s)$ vanishes at $s=0$, as is evident from Eq. (3.9).
Including terms up to $l=3$ the normalization factor can be expressed as
\begin{eqnarray}
\label{5.7}(\hat{\vc{\psi}}^I|\hat{\du{A}}^0|\hat{\vc{\psi}}^I)&=&3|\mu^I_1|^2+\frac{27}{10}|\kappa^I_2|^2+\frac{36}{5}\mathrm{Re}(\kappa^I_2\mu^{I*}_2)+6|\mu^I_2|^2\nonumber\\
&+&\frac{118}{21}|\kappa^I_3|^2+\frac{100}{7}\mathrm{Re}(\kappa^{I*}_3\mu^I_3)+10|\mu^I_3|^2.
\end{eqnarray}
The numerator in Eq. (4.1) at $s=0$ can be expressed similarly as
\begin{equation}
\label{5.8}(\hat{\vc{\psi}}^I|\hat{\du{B}}^I(0)|\hat{\vc{\psi}}^I)=\frac{6}{35}\mathrm{Im}[7\mu^{I*}_1\kappa^I_2+35\mu^{I*}_1\mu^I_2
+6\kappa^I_2\kappa^{I*}_3+30\kappa^{I*}_2\mu^I_3+30\mu^{I*}_2\kappa^I_3+70\mu^{I*}_2\mu^I_3].
\end{equation}
The behavior at large $s$ can be expressed as
\begin{eqnarray}
\label{5.9}(\hat{\vc{\psi}}^I|\hat{\du{B}}^I_S(s)|\hat{\vc{\psi}}^I)&=&Ps+Q_S+O(1/s),\nonumber\\
(\hat{\vc{\psi}}^I|\hat{\du{B}}^I_B(s)|\hat{\vc{\psi}}^I)&=&-Ps+Q_B+O(1/s),
\end{eqnarray}
showing that the terms linear in $s$ precisely cancel. The constant term $Q_S$ is given by
\begin{equation}
\label{5.10}Q_S=\frac{2}{35}\mathrm{Im}[35\mu^{I*}_1\kappa^I_2+105\mu^{I*}_1\mu^I_2+54\kappa^{I*}_2\kappa^I_3+102\kappa^{I*}_2\mu^I_3+114\mu^{I*}_2\kappa^I_3+210\mu^{I*}_2\mu^I_3],
\end{equation}
and the term $Q_B$ is given by
\begin{equation}
\label{5.11}Q_B=-\frac{8}{35}\mathrm{Im}[14\mu^{I*}_1\kappa^I_2+41\kappa^{I*}_2\kappa^I_3+33\kappa^{I*}_2\mu^I_3+14\mu^{I*}_2\kappa^I_3]+\frac{16}{35}\mathrm{Re}(\kappa^{I*}_2\kappa^I_3),
\end{equation}
The sum is
\begin{eqnarray}
\label{5.12}Q_S+Q_B&=&\frac{2}{35}\mathrm{Im}[-21\mu^{I*}_1\kappa^I_2+105\mu^{I*}_1\mu^I_2-110\kappa^{I*}_2\kappa^{I*}_3-30\kappa^{I*}_2\mu^I_3+58\mu^{I*}_2\kappa^I_3+210\mu^{I*}_2\mu^I_3]\nonumber\\
&+&\frac{16}{35}\mathrm{Re}(\kappa^{I*}_2\kappa^I_3).
\end{eqnarray}
It is clear that this expression can have either sign, depending on the coefficients, and that the sign is independent of that in the expression Eq. (5.8). One can check that the velocity
\begin{equation}
\label{5.13}U_{red}(\infty)=\frac{Q_S+Q_B}{(\hat{\vc{\psi}}^I|\hat{\du{A}}^0|\hat{\vc{\psi}}^I)},
\end{equation}
in our examples takes the values given in Sec. IV. It is easy to extend the expressions in Eqs. (5.7) and (5.12) to include higher order mode coefficients.

The two expressions on the left in Eq. (5.9) originate in the mean convective flow generated directly by the surface distortions of the sphere, and the reactive flow generated by the Reynolds force density, respectively. Apparently there is a delicate balance between these two quantities. It follows from Eq. (3.9) that the bulk matrix $\hat{\du{B}}^I_{13B}(s)$ has vanishing $\mu\mu$-elements. Although the matrix originates in the Reynolds stress of the bulk flow, there is a contribution only from the boundary layer of thickness $a/s$. In the limit of large $s$ the boundary layer becomes very thin, but its effect on the sphere velocity depends on the details of its structure, as evident from Eq. (5.12).

Spelman and Lauga \cite{12} studied the boundary layer problem for an active particle in the inertia-dominated limit by the method of matched asymptotic expansions. For the case of radial surface displacements they find a quite complicated expression for the swimming speed involving Gaunt coefficients, which is not easy to compare with our result Eq. (5.13) for $U_{red}(\infty)$. They also find the possibility of sign changes depending on the choice of mode combinations.

\section{\label{VI}Mean Reynolds force density}

It is of interest to consider besides the mean swimming velocity also the net flow pattern of a swimmer. One of us studied the net flow pattern of simple swimmers in the Stokes limit \cite{20}. Here we extend the analysis to arbitrary values of the scale number $s$. Both the direct contribution from the second order surface velocity and the contribution from the second order Reynolds stress depend on the scale number $s$. For the mean second order flow we need to solve a steady state Stokes problem \cite{6}. The direct contribution to the net flow can be evaluated fairly straightforwardly as the solution of the Stokes problem with boundary condition given by the mean second order surface velocity, as given by Eq. (2.9).

In order to find the contribution from the Reynolds stress we must solve the inhomogeneous Stokes problem Eq. (2.8) with driving term given by the mean Reynolds force density
\begin{equation}
\label{6.1}\overline{\vc{f}^{(2)}_R}(\vc{r})=-\frac{\rho}{T}\int^T_0\vc{v}^{(1)}(\vc{r},t)\cdot\nabla\vc{v}^{(1)}(\vc{r},t)\;dt,
\end{equation}
where the time-average on the right is over a period $T$. The resulting mean flow velocity can be expressed as
\begin{equation}
\label{6.2}\overline{\vc{v}^{(2)}_R}(\vc{r})=\frac{1}{\eta}\int_{r'>a}\vc{G}(\vc{r},\vc{r}')\cdot\overline{\vc{f}^{(2)}_R}(\vc{r}')\;d\vc{r}',
\end{equation}
where $\vc{G}(\vc{r},\vc{r}')$ is the Green function of the Stokes equations with no-slip boundary condition on the surface of a fixed sphere with radius $a$ centered at the origin. We shall discuss the Green function in the next section.

In our axisymmetric problem the mean Reynolds force density can be expanded in vector spherical harmonics $\vc{A}_l$ and $\vc{B}_l$,
\begin{equation}
\label{6.3}\overline{\vc{f}^{(2)}_R}(\vc{r})=\sum_{l=1}^\infty\big[f_{Al}(r)\vc{A}_l+f_{Bl}(r)\vc{B}_l\big].
\end{equation}
As before it is convenient to use complex notation. We therefore express the time-average in Eq. (6.1) as
\begin{equation}
\label{6.4}\overline{\vc{f}^{(2)}_R}(\vc{r})=-\frac{1}{2}\rho\;\mathrm{Re}[\vc{v}^*_\omega\cdot\nabla\vc{v}_\omega].
\end{equation}
With the expansion corresponding to Eq. (2.7),
\begin{equation}
\label{6.5}\vc{v}_\omega(\vc{r})=-\omega a\sum^\infty_{l=1}[\kappa_l\vc{v}_l(\vc{r},\alpha)+\mu_l\vc{u}_l(\vc{r})],
\end{equation}
and the understanding that $\kappa_1=0$, the complex function $\vc{v}^*_\omega\cdot\nabla\vc{v}_\omega$ can be expressed as
\begin{equation}
\label{6.6}\vc{v}^*_\omega\cdot\nabla\vc{v}_\omega=\sum_{l=1}^\infty(\hat{\vc{\psi}}|\du{F}_{Al}(r,s)|\hat{\vc{\psi}})\vc{A}_l
+\sum_{l=1}^\infty(\hat{\vc{\psi}}|\du{F}_{Bl}(r,s)|\hat{\vc{\psi}})\vc{B}_l,
\end{equation}
where the elements of the matrices $\du{F}_{Al}(r,s)$ and $\du{F}_{Bl}(r,s)$ can be evaluated from the expansions of the bilinear expressions
\begin{equation}
\label{6.7}\vc{v}^*_j\cdot\nabla\vc{v}_k,\qquad\vc{v}^*_j\cdot\nabla\vc{u}_k,\qquad
\vc{u}^*_j\cdot\nabla\vc{v}_k,\qquad\vc{u}^*_j\cdot\nabla{\vc{u}}_k
\end{equation}
in terms of vector spherical harmonics $\vc{A}_l$ and $\vc{B}_l$. These can be evaluated by use of the orthonormality relations for the $\vc{A}_l$ and $\vc{B}_l$ which read
\begin{eqnarray}
\label{6.8}\int^\pi_0\vc{A}_k\cdot\vc{A}_l\sin\theta\;d\theta&=&2k\delta_{kl},\qquad\int^\pi_0\vc{A}_k\cdot\vc{B}_l\sin\theta\;d\theta=0,\nonumber\\
\int^\pi_0\vc{B}_k\cdot\vc{B}_l\sin\theta\;d\theta&=&(2k+2)\delta_{kl}.
\end{eqnarray}
The expression for the tensor $\nabla\vc{v}$ in spherical coordinates is given by Happel and Brenner \cite{3}.

Finally we can rewrite
\begin{eqnarray}
\label{6.9}(\hat{\vc{\psi}}|\du{F}_{Al}(r,s)|\hat{\vc{\psi}})&=&(\hat{\vc{\psi}}^I|\du{F}^I_{Al}(r,s)|\hat{\vc{\psi}}^I),\nonumber\\
(\hat{\vc{\psi}}|\du{F}_{Bl}(r,s)|\hat{\vc{\psi}})&=&(\hat{\vc{\psi}}^I|\du{F}^I_{Bl}(r,s)|\hat{\vc{\psi}}^I),
\end{eqnarray}
by use of the transformation matrix $\du{T}$ as in Eq. (3.4). A comparison of Eqs. (6.3) and (6.6) yields expressions for the radial functions $f_{Al}(r)$ and $f_{Bl}(r)$. If the coefficient vector $\hat{\vc{\psi}}$ is truncated at low order, then only correspondingly low order matrices $\du{F}_{Al}(r,s)$ and $\du{F}_{Bl}(r,s)$ need to be calculated.

\section{\label{VII}Net flow pattern}

The mean second order flow velocity $\overline{\vc{v}^{(2)}(\vc{r})}$ is defined in the volume $r>a$. It tends to $-\overline{U^{(2)}}\vc{e}_z$ at infinity \cite{6}. The net flow pattern is defined as \cite{20}
\begin{equation}
\label{7.1}\vc{v}'(\vc{r})=\overline{\vc{v}^{(2)}(\vc{r})}+\overline{U^{(2)}}\vc{e}_z.
\end{equation}
This tends to zero at infinity. The flow can be decomposed as a sum of a surface and a bulk contribution,
\begin{equation}
\label{7.2}\vc{v}'(\vc{r})=\vc{v}'_S(\vc{r})+\vc{v}'_V(\vc{r}),
\end{equation}
where each term has a decomposition as in Eq. (7.1),
\begin{eqnarray}
\label{7.3}\vc{v}_S'(\vc{r})&=&\overline{\vc{v}^{(2)}_S(\vc{r})}+\overline{U_{2S}}\vc{e}_z,\nonumber\\
\vc{v}'_V(\vc{r})&=&\overline{\vc{v}^{(2)}_V(\vc{r})}+\overline{U_{2B}}\vc{e}_z.
\end{eqnarray}
Stokes flows corresponding to a moving sphere with velocities $\overline{U_{2S}}$ and $\overline{U_{2B}}$ respectively are included such that each of these flow patterns decays to zero faster than $1/r$ at infinity, corresponding to vanishing net force. The mean second order flow velocities can be expressed as
\begin{eqnarray}
\label{7.4}\overline{\vc{v}^{(2)}_S(\vc{r})}&=&\overline{\vc{v}}_{2S}(\vc{r})-\overline{U_{2S}}\vc{e}_z+\vc{v}^{St}_S(\vc{r}),\nonumber\\
\overline{\vc{v}^{(2)}_V(\vc{r})}&=&\overline{\vc{v}^{(2)}_R}(\vc{r})-\overline{U_{2B}}\vc{e}_z+\vc{v}^{St}_B(\vc{r}),
\end{eqnarray}
where $\overline{\vc{v}}_{2S}(\vc{r})$ is the solution of the homogeneous Stokes equations which equals the second order surface velocity $\overline{\vc{u}}_S(\vc{s})$ on $r=a$ and tends to zero at infinity, and $\overline{\vc{v}^{(2)}_R}(\vc{r})$ is given by Eq. (6.2). The fact that this works with the velocities $\overline{U_{2S}}$ and $\overline{U_{2B}}$ as calculated above in Secs. III and V proves that the calculation is consistent.

Each pattern can be decomposed into vector spherical harmonics as in Eq. (6.3),
\begin{eqnarray}
\label{7.5}\overline{\vc{v}}_{2S}(\vc{r})&=&\sum_{l=1}^\infty\big[v_{SAl}(r)\vc{A}_l+v_{SBl}(r)\vc{B}_l\big],\nonumber\\
\overline{\vc{v}^{(2)}_R}(\vc{r})&=&\sum_{l=1}^\infty\big[v_{VAl}(r)\vc{A}_l+v_{VBl}(r)\vc{B}_l\big].
\end{eqnarray}
The flow $\overline{\vc{v}}_{2S}(\vc{r})$ satisfies the homogeneous Stokes equations, so that the radial functions $v_{SAl}(r)$ and $v_{SBl}(r)$ follow from the expressions for the mode functions $\vc{u}_l(\vc{r})$ and $\vc{v}^0_l(\vc{r})$ given in Eqs. (2.5) and (3.1), with coefficients corresponding to the mean second order surface velocity at $r=a$. The functions $v_{VAl}(r)$ and $v_{VBl}(r)$ must be derived from the integral expression in Eq. (6.2).

In order to derive the second expansion in Eq. (7.5) we need the corresponding expansion of the Green function. This can be found as an extension of the antenna theorems derived by Schmitz and Felderhof \cite{21}. The explicit expression at angular order $l$ reads
\begin{eqnarray}
\label{7.6}
G_{<lAA}(r,b)&=&\frac{l+1}{4l^2-1}\;\frac{r^{2l-1}-a^{2l-1}}{b^{l-2}r^l},\nonumber\\
G_{>lAA}(r,b)&=&\frac{l+1}{4l^2-1}\;\frac{b^{2l-1}-a^{2l-1}}{b^{l-2}r^l},\nonumber\\
G_{<lBA}(r,b)&=&\frac{l}{4l+2}\;a^{2l-1}\;\frac{r^2-a^2}{b^{l-2}r^{l+2}},\nonumber\\
G_{>lBA}(r,b)&=&\frac{l}{4l+2}\;\frac{b^{2l+1}-b^{2l-1}r^2+a^{2l-1}r^2-a^{2l+1}}{b^{l-2}r^{l+2}},\nonumber\\
G_{<lAB}(r,b)&=&\frac{l+1}{4l+2}\;\frac{r^{2l+1}-r^{2l-1}b^2+a^{2l-1}b^2-a^{2l+1}}{b^{l}r^{l}},\nonumber\\
G_{>lAB}(r,b)&=&\frac{l+1}{4l+2}\;a^{2l-1}\;\frac{b^2-a^2}{b^lr^l},\nonumber\\
G_{<lBB}(r,b)&=&\frac{l}{4 (2l+1)(2l+3)}\;\frac{4r^{2l+3}+(4l^2+4l-3)a^{2l-1}(a^2b^2+a^2r^2-b^2r^2)-(2l+1)^2 a^{2l+3}}{b^lr^{l+2}},\nonumber\\
G_{>lBB}(r,b)&=&\frac{l}{4 (2l+1)(2l+3)}\;\frac{4b^{2l+3}+(4l^2+4l-3)a^{2l-1}(a^2b^2+a^2r^2-b^2r^2)-(2l+1)^2 a^{2l+3}}{b^lr^{l+2}},\nonumber\\
\end{eqnarray}
where the functions $G_{<l\alpha\beta}$ apply for $a<r<b$ and the functions $G_{>l\alpha\beta}$ apply for $r>b$. The functions are continuous at $r=b$ and reduce to the Schmitz-Felderhof antenna theorems in the limit $a\rightarrow 0$. The functions $v_{VAl}(r)$ and $v_{VBl}(r)$ are given by radial integrals in terms of the functions $f_{Al}(r)$ and $f_{Bl}(r)$,
\begin{eqnarray}
\label{7.7}
v_{VAl}(r)&=&\frac{1}{\eta}\int^\infty_r\big[G_{<lAA}(r,b)f_{Al}(b)+G_{<lAB}(r,b)f_{Bl}(b)\big]\;db\nonumber\\
&+&\frac{1}{\eta}\int^r_a\big[G_{>lAA}(r,b)f_{Al}(b)+G_{>lAB}(r,b)f_{Bl}(b)\big]\;db,\nonumber\\
v_{VBl}(r)&=&\frac{1}{\eta}\int^\infty_r\big[G_{<lBA}(r,b)f_{Al}(b)+G_{<lBB}(r,b)f_{Bl}(b)\big]\;db\nonumber\\
&+&\frac{1}{\eta}\int^r_a\big[G_{>lBA}(r,b)f_{Al}(b)+G_{>lBB}(r,b)f_{Bl}(b)\big]\;db.
\end{eqnarray}
For the functions $f_{Al}(b)$ and $f_{Bl}(b)$ which occur these integrals can be performed in analytic form.

\section{\label{VIII}Stokes stream matrices}

For the axisymmetric problem the net flow pattern can be derived from a Stokes stream function. This is useful for the plotting of streamlines. In the present section we derive explicit results for the steam function for low order swimmers with stroke truncated at angular number $L=2$. The reults can be summarized in matrix form with three-dimensional Stokes stream matrices in the representation of modes $\vc{u}_1,\vc{v}^0_2,\vc{u}_2$, as given by Eqs. (2.5) and (3.1). The total Stokes stream matrix is the sum of a surface part and a volume part,
\begin{equation}
\label{8.1}\vc{\Psi}^I_{12}(r,\theta)=\vc{\Psi}^I_{12S}(r,\theta)+\vc{\Psi}^I_{12V}(r,\theta),
\end{equation}
with the surface part calculated from the steady state Stokes equations with boundary condition given by the second order surface velocity and with the volume part calculated from Eq. (6.2). These matrices are calculated in a frame with the fluid at rest at infinity. The net Stokes stream functions are given by
\begin{eqnarray}
\label{8.2}\vc{\Psi}^{I\prime}_{12S}=\vc{\Psi}^I_{12S}+\frac{1}{2}\omega a\du{B}^I_{12S}\psi^{St},\nonumber\\
\vc{\Psi}^{I\prime}_{12V}=\vc{\Psi}^I_{12V}+\frac{1}{2}\omega a\du{B}^I_{12B}\psi^{St},
\end{eqnarray}
where $\psi^{St}(r,\theta)$ is the Stokes stream function for a sphere of radius $a$ with no-slip boundary condition and moving with unit velocity in the positive $z$ direction, given explicitly by \cite {22}
\begin{equation}
\label{8.3}\psi^{St}(r,\theta)=\frac{1}{4}\bigg(3ar-\frac{a^3}{r}\bigg)\sin^2\theta.
\end{equation}
The second term on the right in Eq. (8.2) must be added to make sure that the net flow pattern corresponds to a swimmer exerting no net force on the fluid \cite{6}. The matrix $\vc{\Psi}^I_{12V}$ vanishes for $r=a$.

The calculation of the matrices $\vc{\Psi}^I_{12S}$ and $\vc{\Psi}^I_{12V}$ is complicated, but can be performed in analytic form. The matrix $\vc{\Psi}^I_{12S}$ has the structure
\begin{equation}
\label{8.4}\vc{\Psi}^I_{12S}=\omega a^3\sin^2\theta\left(\begin{array}{ccc}0&S_{12}&S_{13}
\\S_{12}^*&S_{22}&S_{23}
\\S_{13}^*&S_{23}^*&0\end{array}\right),
\end{equation}
The element $S_{22}$ is real. It is convenient to introduce the shorthand notation
\begin{equation}
\label{8.5}p=(1+i)s,\qquad m=(1-i)s.
\end{equation}
With this notation the element $S_{12}$ is given by
\begin{eqnarray}
\label{8.6}S_{12}=\frac{3i}{160(1+m)ar^3}&\bigg[&3(9+9m+4m^2)a^4-(29+29m+4m^2)a^2r^2+4 (3+3m-2m^2)r^4\nonumber\\
&+&5\bigg((9+9m+4m^2)a^4-(3+3m+4m^2)a^2r^2\bigg)\cos2\theta\bigg].
\end{eqnarray}
The element $S_{13}$ is given by
\begin{equation}
\label{8.7}S_{13}=\frac{-3i}{32ar^3}\bigg[3(a^4+5a^2r^2-4r^4)+(5a^4-3a^2r^2)\cos2\theta\bigg].
\end{equation}
The element $S_{22}$ is given by
\begin{equation}
\label{8.8}S_{22}=\frac{9s^2(1+s)(a^2-r^2)}{56(1+2s+2s^2)r^4}\bigg[(3a^2+4r^2)\cos\theta+21a^2\cos\theta\cos2\theta\bigg].
\end{equation}
The element $S_{23}$ is given by
\begin{eqnarray}
\label{8.9}S_{23}=\frac{-9i}{224(1+p)r^4}&\bigg[&\bigg(3(3+3p+p^2)a^4+(17+17p+p^2)a^2r^2-4(3+3p+p^2)r^4\bigg)\cos\theta\nonumber\\
&+&21\bigg((3+3p+p^2)a^4-(1+p+p^2)a^2r^2\bigg)\cos\theta\cos2\theta\bigg].
\end{eqnarray}

The matrix $\vc{\Psi}^I_{12V}$ has the structure
\begin{equation}
\label{8.10}\vc{\Psi}^I_{12V}=\omega a^3\sin^2\theta\left(\begin{array}{ccc}0&V_{12}&0
\\V_{12}^*&V_{22}&V_{23}
\\0&V_{23}^*&0\end{array}\right),
\end{equation}
The elements which vanish do so on account of a general theorem which we proved earlier \cite{6}. The element $V_{22}$ is real. The matrix elements in Eq. (8.10) can be evaluated in analytic form. The complicated expressions are listed in the Appendix. In the next section we show net flow patterns calculated on the basis of these expressions.

\section{\label{IX}Flow pattern calculations}

Explicit calculations can be performed for the simple swimmers studied in Sec. IV. Each of these is characterized by a coefficient vector $\hat{\vc{\psi}}^I$ specifying the first order stroke of the swimmer. For the swimmers of Sec. IV the vector consists of two, three or five complex coefficients. Much of the calculation can be performed in terms of corresponding two-, three- or five-dimensional matrices, which can be calculated once and for all without specifying the swimmer. The three-dimensional Stokes stream matrices are given explicitly in the preceding section and in the Appendix. In general the matrices are complex and depend on scale number $s$ and the dimensionless ratio $r/a$. We showed in our earlier work \cite{6},\cite{8} that the theory of swimming simplifies considerably if the first order flow is irrotational. In that case the mean Reynolds flow velocity $\overline{\vc{v}^{(2)}_R}(\vc{r})$ in Eq. (6.2) vanishes, and the mean swimming velocity and the net flow velocity are independent of $s$.

We consider first the simplest potential swimmer for which the first order irrotational flow field is a superposition of a dipolar and a quadrupolar flow field, corresponding to moments $\mu^I_1=1$, $\mu^I_2=i/\sqrt{2}$, and all other moments vanishing. In Sec. IV we calculated for this swimmer the reduced swimming velocity $U_{red}=1/\sqrt{2}$, independent of $s$. With amplitude factor $\varepsilon$ the surface displacement is
\begin{equation}
\label{9.1}\vc{\xi}(\vc{s},t)=\varepsilon a\big[\vc{B}_1(\theta)\sin(\omega t)-\frac{1}{\sqrt{2}}\vc{B}_2(\theta)\cos(\omega t)\big],
\end{equation}
with vector spherical harmonics given in Eq. (2.6). The corresponding first order flow is
\begin{eqnarray}
\label{9.2}\vc{v}^{(1)}(\vc{r},t)&=&\varepsilon a\omega\bigg[\frac{a^3}{r^3}\vc{B}_1(\theta)\cos(\omega t)+\frac{a^4}{\sqrt{2}r^4}\vc{B}_2(\theta)\sin(\omega t)\bigg],\nonumber\\
p^{(1)}(\vc{r},t)&=&\varepsilon\rho a^2\omega^2\bigg[\frac{a^2}{r^2}P_1(\cos\theta)\sin(\omega t)-\frac{a^3}{\sqrt{2}r^3}P_2(\cos\theta)\cos(\omega t)\bigg].
\end{eqnarray}
The first order pressure deviation follows from Eq. (2.5). Since the first order flow velocity is irrotational the mean Reynolds force density may be expressed as the gradient of a scalar function which may be identified with a second order mean pressure deviation proportional to the mass density \cite{6}. As a consequence the Reynolds flow velocity $\overline{\vc{v}^{(2)}_R}(\vc{r})$, given by Eq. (6.2), vanishes. The mean second order flow velocity has only a surface contribution, and is independent of $s$. The relevant elements of the matrices $\vc{\Psi}^I_{12V}$ and $\du{B}^I_{12B}$ in Eq. (8.2) vanish, and it suffices to consider the two-dimensional matrices
\begin{equation}
\label{9.3}\vc{\Psi}^I_{12S}=\omega a^3\sin^2\theta\left(\begin{array}{cc}0&S_{13}
\\S_{13}^*&0\end{array}\right),\qquad\du{B}^I_{12S}=\left(\begin{array}{cc}0&-3i
\\3i&0\end{array}\right),
\end{equation}
with $S_{13}$ given by Eq. (8.7). From the matrix $\du{B}^I_{12S}$ we find for the second order mean swimming velocity
\begin{equation}
\label{9.4}\overline{U_2}=\overline{U_{2S}}=\frac{1}{2}\varepsilon^2a\omega(1,-\frac{i}{\sqrt{2}}).\du{B}^I_{12S}.(1,\frac{i}{\sqrt{2}})=\frac{3}{\sqrt{2}}\;\varepsilon^2a\omega.
\end{equation}
From the matrix $\vc{\Psi}^I_{12S}$ we find in the same way the stream function
\begin{equation}
\label{9.5}\psi_{12S}(r,\theta)=\frac{1}{16}\overline{U_2}\frac{a}{r^3}\bigg[3(a^4+5a^2r^2-4r^4)+(5a^4-3a^2r^2)\cos2\theta\bigg]\sin^2\theta.
\end{equation}
 In the notation of Ref. 6 the corresponding flow velocity is given by
\begin{equation}
\label{9.6}\overline{\vc{v}}_{2S}(\vc{r})=\overline{U_2}\bigg[-\frac{a}{r}\vc{A}_1-\frac{21a^3-5ar^2}{20r^3}\vc{B}_1-\frac{2a^3}{35r^3}\vc{A}_3-\frac{7a^5-3a^3r^2}{28r^5}\vc{B}_3\bigg],
\end{equation}
corresponding to surface velocity
\begin{equation}
\label{9.7}\overline{\vc{u}}_S(\vc{s})=\overline{U_2}\bigg[-\vc{A}_1-\frac{4}{5}\vc{B}_1-\frac{2}{35}\vc{A}_3-\frac{1}{7}\vc{B}_3\bigg].
\end{equation}
In the derivation of Eq. (9.6) we have used the orthonormality relations Eq. (6.8).
We recall that $\vc{A}_1=\vc{e}_z$ and $\vc{B}_1=\vc{e}_z-3\vc{e}_r\cos\theta$. The surface average of $\vc{B}_1,\vc{A}_3,\vc{B}_3$ vanishes. The first term in Eq. (9.7) is in agreement with a general theorem \cite{7}. The surface velocity $\overline{\vc{u}}_S(\vc{s})$ is the primary quantity which can be calculated from the surface displacement in Eq. (9.1) and the first order flow velocity given by Eq. (9.2). The second order flow velocity in Eq. (9.6) is the corresponding solution of the steady state Stokes equations which tends to zero at infinity.

The Stokes flow in Eq. (7.4) is given by
\begin{equation}
\label{9.8}\vc{v}^{St}_S(\vc{r})=\overline{U_2}\bigg[\frac{a}{r}\big(\vc{A}_1-\frac{1}{4}\vc{B}_1\big)+\frac{a^3}{4r^3}\vc{B}_1\bigg].
\end{equation}
This must be added to Eq. (9.6) in order to cancel the $a/r$ term. The latter corresponds to an Oseen flow generated by a force acting on the fluid. In swimming there is no net force acting.
From the sum of Eqs. (9.6) and (9.8) we find for the net flow velocity
\begin{equation}
\label{9.9}\vc{v}^{\prime}_{12}(r,\theta)=
\overline{U_2}\bigg[-\frac{4a^3}{5r^3}\vc{B}_1-\frac{2a^3}{35r^3}\vc{A}_3-\frac{7a^5-3a^3r^2}{28r^5}\vc{B}_3\bigg],
\end{equation}
corresponding to superposition of a dipolar and an octupolar flow. The latter has nonvanishing vorticity directed in the azimuthal direction,
\begin{equation}
\label{9.10}\nabla\times\vc{v}^{\prime}_{12}=-\frac{3}{16}\;\overline{U_2}\;\frac{a^3}{r^4}\big[\sin\theta+5\sin3\theta\big]\vc{e}_\varphi.
\end{equation}
The vorticity is generated by the no-slip boundary condition at the undulating surface. Taking the Laplacian of Eq. (9.9) we obtain a contribution to the mean pressure deviation proportional to the shear viscosity.

From Eqs. (8.3) and (9.3)  we find for the stream function of the net flow
\begin{equation}
\label{9.11}\psi^{\prime}_{12}(r,\theta)=\frac{1}{16}\;\overline{U_2}\;\frac{a^3}{r^3}\big[3a^2+11r^2+(5a^2-3r^2)\cos2\theta\big]\sin^2\theta.
\end{equation}
This yields the streamlines of the net flow, as shown in Fig. 3. The net flow pattern is long range, falling off as $1/r^3$ at large distance. In terms of the modes defined in Eqs. (2.5) and (3.1) the flow is given by
\begin{equation}
\label{9.12}\vc{v}^{\prime}_{12}(r,\theta)=\overline{U_2}\bigg[\frac{4}{5}\vc{u}_1+\frac{1}{4}\vc{u}_3-\frac{3}{20}\vc{v}^0_3\bigg],
\end{equation}
showing two potential modes and one vortex mode. The flow velocity is independent of mass density and shear viscosity of the fluid. Only the vortex mode $\vc{v}^0_3$ contributes to the vorticity in Eq. (9.10). The mode can be viewed as being composed of three vortex rings, two of the same vorticity front and aft of the sphere, and a central one of opposite vorticity. The relative contributions to the total vorticity are given by weights $(0.313,0.374,0.313)$.

In the following we present numerical results for the net time-averaged flow pattern of several selected swimmers with nonvanishing Reynolds flow.
First we consider the swimmer specified by three coefficients $\mu_1^I=1,\;\kappa^I_2=-4i\sqrt{2}/3,\;\mu^I_2=11i/(5\sqrt{2})$. The reduced swimming velocity $U_{red}(s)$ for this swimmer was plotted in Fig. 1. In Fig. 4 we show the streamlines of the net flow, as calculated from the corresponding Stokes stream function, for scale number $s=0.1$. At this small scale number the net flow is dominated by the surface contribution $\vc{v}'_S(\vc{r})$. The reduced swimming velocity is $U_{red}(0.1)=1.179$, and the relative contribution of the bulk and surface terms is $U_B(0.1)/U_S(0.1)=-10^{-5}$. The net flow is quite similar to that calculated in the Stokes limit $s=0$.

In Fig. 5 we show the streamlines of the net flow for the same swimmer with scale number $s=10$. A comparison with Fig. 4 shows a significant distortion of the streamlines. Interestingly the flow shows a detached vortex ring. The reduced swimming velocity is $U_{red}(10)=1.730$, and the relative contribution of the bulk and surface terms is $U_B(10)/U_S(10)=-0.499$. The ratio suggests that there is significant cancellation of the surface and bulk contributions to the net flow. In Fig. 6 we show this for the $z$-component of the net flow velocity in the equatorial plane $\theta=\pi/2$.

In Fig. 7 we show the streamlines of the net flow for the $123$-swimmer with coefficient vector given by
the optimized values of Eq. (4.5). We choose scale number $s=10$, corresponding to the reduced swimming velocity $U_{red}(10)=1.904$, and ratio $U_B(10)/U_S(10)=-0.285$. The flow pattern is similar to that shown in Fig. 5, but there is no detached vortex ring.

In the limit of large $s$ the boundary layer of thickness $a/s$ becomes very thin. Correspondingly the radial force density functions $f_{Al}(r)$ and $f_{Bl}(r)$ in Eq. (7.7) consist of a rapidly varying part at the boundary and a slowly varying part for larger values of $r$. The bulk flow functions $v_{VAl}(r)$ and $v_{VBl}(r)$ decompose correspondingly. The radial dependence of these functions is determined by that of the Green function in Eq. (7.6). The final flow pattern is a superposition of many contributions which can be evaluated for any chosen particular stroke. The main qualitative insight is that the net flow $\vc{v}'_V(\vc{r})$ can be neglected for small scale number $s$, and that the net flows $\vc{v}'_S(\vc{r})$ and $\vc{v}'_V(\vc{r})$ nearly cancel for large $s$. In the next section we analyze this behavior in more detail.

\section{\label{X}Analysis of net flow pattern}

The net flow pattern defined in Eq. (7.1) can be written as a sum of two terms as in Eq. (7.2). For the dipole-quadrupole model studied in the preceding section only the surface part contributes. More generally we must consider also the volume contribution $\vc{v}'_V(\vc{r})$. Outside a boundary layer of thickness $\lambda=\sqrt{2\eta/(\omega\rho)}$ both parts are solutions of the linear Stokes equations and can be written as superpositions of Stokes flows $\{\vc{u}_l,\vc{v}^0_l\}$. We define corresponding moments $\{M_l,K_l\}$ from the decomposition
\begin{equation}
\label{10.1}\vc{v}'_{ou}(\vc{r})=\varepsilon^2a \omega\big[M_1\vc{u}_1+K_2\vc{v}^0_2+M_2\vc{u}_2+K_3\vc{v}^0_3+M_3\vc{u}_3+...\big],
\end{equation}
where the subscript $ou$ indicates that only the part outside the boundary layer is being considered.
For the dipole-quadrupole model we found in Eq. (9.12) for the first few moments
\begin{equation}
\label{10.2}(M_1,K_2,M_2,K_3,M_3)=\frac{3}{\sqrt{2}}\;\big(\frac{4}{5},0,0,-\frac{3}{20},\frac{1}{4}\big),
\end{equation}
and all higher order moments vanishing. In that case there is no boundary layer. More generally we find a set of moments which depend on the scale number $s=a/\lambda$.

For general stroke $\hat{\vc{\psi}}^I$ the net flow can be evaluated from the Stokes matrices $\vc{\Psi}^{I\prime}_S$ and $\vc{\Psi}^{I\prime}_V$, as defined in Eq. (8.2). We consider first the matrix $\vc{\Psi}^{I\prime}_S$.
By use of the identities \cite{3}
\begin{eqnarray}
\label{10.3}\sin^2\theta&=&2\mathcal{J}_2(\cos\theta),\qquad\cos\theta\sin^2\theta=2\mathcal{J}_3(\cos\theta),\nonumber\\
\cos2\theta\sin^2\theta&=&-\frac{6}{5}\mathcal{J}_2(\cos\theta)+\frac{16}{5}\mathcal{J}_4(\cos\theta),\nonumber\\
\cos\theta\cos2\theta\sin^2\theta&=&-\frac{2}{7}\mathcal{J}_3(\cos\theta)+\frac{16}{7}\mathcal{J}_5(\cos\theta),\nonumber\\
\cos3\theta\sin^2\theta&=&-\frac{18}{7}\mathcal{J}_3(\cos\theta)+\frac{32}{7}\mathcal{J}_5(\cos\theta),
\end{eqnarray}
the expressions in Eqs. (8.6-9) can be turned into linear combinations of Gegenbauer functions $\mathcal{I}_l(\cos\theta)$. We note the relations
\begin{equation}
\label{10.4}\vc{S}_{op}\;\frac{a^{l+2}\mathcal{J}_{l+1}(\cos\theta)}{r^l}=\frac{1}{l+1}\;\vc{u}_l,\qquad \vc{S}_{op}\;\frac{a^l\mathcal{J}_{l+1}(\cos\theta)}{r^{l-2}}=\frac{1}{l+1}\;\vc{v}^0_l,
\end{equation}
where $\vc{S}_{op}$ is the linear Stokes operator defined by the rule $\vc{v}=\vc{S}_{op}\psi$ with \cite{22}
\begin{equation}
\label{10.5}v_r(r,\theta)=\frac{1}{r^2\sin\theta}\frac{\partial\psi}{\partial\theta},\qquad v_\theta(r,\theta)=\frac{-1}{r\sin\theta}\frac{\partial\psi}{\partial r},\qquad v_\varphi=0.
\end{equation}
Correspondingly the matrix  $\vc{\Psi}^{I\prime}_S$ can be decomposed into matrices $\vc{\mathcal{M}}_{Sl},\vc{\mathcal{K}}_{Sl}$ as
\begin{equation}
\label{10.6}\vc{\Psi}^{I\prime}_S(r,\theta)=\omega a^3\sum^\infty_{l=1}(l+1)\bigg[\frac{a^l}{r^l}\vc{\mathcal{M}}_{Sl}+\frac{a^{l-2}}{r^{l-2}}\vc{\mathcal{K}}_{Sl}\bigg]\mathcal{J}_{l+1}(\cos\theta).
\end{equation}
For stroke $\hat{\vc{\psi}}^I$ the surface contribution to the moments in Eq. (10.1) is then given by
\begin{equation}
\label{10.7}M_{Sl}=(\hat{\vc{\psi}}^I|\vc{\mathcal{M}}_{Sl}|\hat{\vc{\psi}}^I),\qquad K_{Sl}=(\hat{\vc{\psi}}^I|\vc{\mathcal{K}}_{Sl}|\hat{\vc{\psi}}^I).
\end{equation}

From the element $S'_{12}$ of $\vc{\Psi}^{I\prime}_S$ we find the nonvanishing elements
\begin{eqnarray}
\label{10.8}\mathcal{M}_{S1;12}&=&\frac{i}{10}\;\frac{-3-3m+m^2}{1+m},\qquad
\mathcal{K}_{S3;12}=-\frac{3i}{40}\;\frac{3+3m+4m^2}{1+m},\nonumber\\
\mathcal{M}_{S3;12}&=&\frac{3i}{40}\;\frac{9+9m+4m^2}{1+m}.
\end{eqnarray}
From the element $S'_{13}$ we find the nonvanishing elements
\begin{equation}
\label{10.9}\mathcal{M}_{S1;13}=-\frac{6i}{5}\qquad \mathcal{K}_{S3;13}=\frac{9i}{40}\qquad
\mathcal{M}_{S3;13}=-\frac{3i}{8}.
\end{equation}
From the element $S'_{22}$ we find the nonvanishing elements
\begin{eqnarray}
\label{10.10}\mathcal{K}_{S2;22}&=&-\frac{3}{7}\;\frac{s^2(1+s)}{1+2s+2s^2},\qquad
\mathcal{M}_{S2;22}=\frac{3}{7}\;\frac{s^2(1+s)}{1+2s+2s^2},\nonumber\\
\mathcal{K}_{S4;22}&=&-\frac{54}{35}\;\frac{s^2(1+s)}{1+2s+2s^2},\qquad
\mathcal{M}_{S4;22}=\frac{54}{35}\;\frac{s^2(1+s)}{1+2s+2s^2}.
\end{eqnarray}
From the element $S'_{23}$ we find the nonvanishing elements
\begin{eqnarray}
\label{10.11}
\mathcal{K}_{S2;23}=\frac{3i}{28}\;\frac{3+3p+p^2}{1+p},\qquad
\mathcal{M}_{S2;23}&=&-\frac{3i}{28}\;\frac{5+5p+p^2}{1+p},\nonumber\\
\mathcal{K}_{S4;23}=\frac{27i}{70}\;\frac{1+p+p^2}{1+p},\qquad
\mathcal{M}_{S4;23}&=&-\frac{27i}{70}\;\frac{3+3p+p^2}{1+p}.
\end{eqnarray}
The $21$-, $31$- and $32$- elements are found by complex conjugation.
For stroke $\hat{\vc{\psi}}^I=(1,0,i/\sqrt{2})$ we recover the moments in Eq. (10.2) by use of Eq. (10.7).

It is evident from the above expressions that the listed elements $\{\mathcal{K}_{lS;jk},\mathcal{M}_{lS;jk}\}$ tend to finite values in the Stokes limit $s=0$. For large scale number $s$ the elements increase linearly with $s$. We show below that the limiting behavior for large $s$ is canceled by corresponding behavior of the elements $\{\mathcal{K}_{lV;jk},\mathcal{M}_{lV;jk}\}$. The sum of both sets of elements tends to a finite limit as $s\rightarrow\infty$. The cancelation between both parts is similar to that found in Eq. (5.9).

We define matrices $\vc{\mathcal{M}}_{Vl},\vc{\mathcal{K}}_{Vl}$ from the behavior of the Stokes matrix $\vc{\Psi}^{I\prime}_V$ outside the boundary layer. This part is obtained by omitting terms which decay exponentially with distance $r$. The remaining non-exponential part $\vc{\Psi}^{I\prime}_{Vou}$ can be decomposed as in Eq. (10.6). The explicit expressions can be found from the matrix elements listed in the Appendix. The matrices $\vc{\mathcal{M}}_{Vl},\vc{\mathcal{K}}_{Vl}$ tend to zero as $s^2$ for small $s$. The expressions are rather complicated, and we quote only the asymptotic behavior for large $s$ of the sums $\vc{\mathcal{M}}_{l}=\vc{\mathcal{M}}_{Sl}+\vc{\mathcal{M}}_{Vl}$ and $\vc{\mathcal{K}}_{l}=\vc{\mathcal{K}}_{Sl}+\vc{\mathcal{K}}_{Vl}$.

For large $s$ we find for the $3\times 3$ submatrices in the upper left-hand corner the behavior
\begin{eqnarray}
\label{10.12}\mathcal{M}_{1;12}&=&\mathcal{M}_{1;21}^*=-\frac{3}{10}\;i+O(s^{-1}),\qquad\mathcal{M}_{1;13}=\mathcal{M}_{1;31}^*=-\frac{6i}{5},\nonumber\\
\mathcal{K}_{3;12}&=&\mathcal{K}_{3;21}^*=-\frac{99}{40}\;i+O(s^{-1}),\qquad
\mathcal{M}_{3;12}=\mathcal{M}_{3;21}^*=\frac{93}{40}\;i+O(s^{-1}),\nonumber\\
 \mathcal{K}_{3;13}&=&\mathcal{K}_{3;31}^*=\frac{9i}{40},\qquad
\mathcal{M}_{3;13}=\mathcal{M}_{3;31}^*=-\frac{3i}{8},\nonumber\\
\mathcal{K}_{2;22}&=&\frac{9}{28}+O(s^{-1}),\qquad
\mathcal{M}_{2;22}=-\frac{9}{28}+O(s^{-1}),\nonumber\\
\mathcal{K}_{4;22}&=&\frac{72}{35}+O(s^{-1}),\qquad
\mathcal{M}_{4;22}=-\frac{72}{35}+O(s^{-1}),\nonumber\\
\mathcal{K}_{2;23}&=&\mathcal{K}_{2;32}^*=\frac{6}{7}\;i+O(s^{-1}),\qquad
\mathcal{M}_{2;23}=\mathcal{M}_{2;32}^*=-\frac{6}{7}\;i+O(s^{-1}),\nonumber\\
\mathcal{K}_{4;23}&=&\mathcal{K}_{4;32}^*=\frac{171}{35}\;i+O(s^{-1}),\qquad
\mathcal{M}_{4;23}=\mathcal{M}_{4;32}^*=-\frac{171}{35}\;i+O(s^{-1}).\nonumber\\
\end{eqnarray}
The values at $s=0$ can be found from Eqs. (10.8-11). All other elements of the submatrices vanish.

The expressions show that if in the stroke $\hat{\vc{\psi}}^I$ both coefficients $\mu_1^I$ and $\kappa^I_2$ differ from zero, then the net flow carries a dipole moment $M_1$, a stresslet $K_2$, a quadrupole moment $M_2$, as well as the higher order moments $K_3$ and $M_3$. As an example we consider the stroke given by $(\mu^I_1,\kappa^I_2,\mu^I_2)=(1,-4i\sqrt{2}/3,11i/(5\sqrt{2}))$. In this case we find for the moments in the Stokes limit $s=0$ the values $M_1(0)=46\sqrt{2}/25,\;K_2(0)=0,\;M_2(0)=0,\;K_3(0)=-219/(100\sqrt{2}),\;M_3(0)=21/(4\sqrt{2}),\;K_4(0)=0,\;M_4(0)=0$. In the inertia-dominated limit $s\rightarrow\infty$ the moments become $M_1(\infty)=46\sqrt{2}/25,\;K_2(\infty)=8/7,\;M_2(\infty)=-8/7,\;K_3(\infty)=-1419/(100\sqrt{2}),\;M_3(\infty)=281/(20\sqrt{2}),\;K_4(\infty)=256/35,\;M_4(\infty)=-256/35$. At intermediate values of the scale number the moments vary weakly with $s$, similar to the behavior shown in Figs. 1 and 2.

\section{\label{XI}Discussion}

In our theory the phenomenon of swimming is explained to be a consequence of convection of the body in the flow generated by wave-type distortions of the body shape with fluid flow caused via the no-slip boundary condition. As we showed above, for swimmers characterized by a sizable scale number the net time-averaged flow pattern arises as a delicate balance between the directly generated flow and the indirect flow caused by the Reynolds force density. As a consequence of this balance, also the mean swimming velocity has an interesting dependence on scale number, as shown in Figs. 1 and 2. Rather elaborate calculations are required to find the mean swimming velocity and the net flow patten for given time-periodic stroke.

We discuss briefly the relation of our derivation to the active particle point of view. In the latter the dynamics of the swimming stroke is ignored and only the net flow pattern on a coarse time scale is considered. In the simplest theories the net flow pattern is postulated as a solution of the steady state Stokes equations. By way of illustration we consider the $A_1B_1$-active particle. For this the steady state flow pattern is assumed to take the form
\begin{equation}
\label{11.1}\vc{v}(\vc{r})=-U\vc{e}_z+m_1\frac{a^3}{r^3}(-\vc{I}+3\vc{e}_r\vc{e}_r)\cdot\vc{e}_z=-U\vc{e}_z+m_1\vc{u}_1,
\end{equation}
with unit tensor $\vc{I}$, swimming velocity $U$, and dipole moment $m_1$ in the $z$-direction. The corresponding net flow pattern $\vc{v}'(\vc{r})=\vc{v}(\vc{r})+U\vc{e}_z$ has a dipolar form. In particular for $m_1=\frac{1}{2}U$ Lighthill's coefficient $A_1$ vanishes and $B_1=\frac{3}{2}U$. This $B_1$-active particle is regarded as a squirmer. It is conceivable that such a net flow pattern is generated on average by the wave motion of cilia, provided short distance behavior is neglected. In the work of Refs. 9-12 corrections to the $B_1B_2$-flow pattern due to Reynolds stress are studied by the method of matched asymptotic expansion.

In our derivation we take a more detailed point of view and derive the net flow pattern to second order in amplitude as a time-average of the flow generated by a harmonically varying stroke and as a function of scale number $s$. Even for relatively simple strokes the net flow pattern is rather convoluted and depends in a subtle way on the generating stroke.

The special case where the first order flow velocity is irrotational is of particular interest. In that case the mean swimming velocity and the net flow velocity are independent of mass density and shear viscosity of the fluid. For the simple example shown in Sec. IX the net flow exhibits an interesting vortex structure. The example suggests that vortex shedding is an indispensable feature of swimming.
\appendix

\newpage

\section{\label{A}}
In this Appendix we provide expressions for the matrix elements in Eq. (8.10). The element $V_{12}$ is given by
\begin{equation}
\label{A.1}V_{12}=G_{120}\Gamma[0,m]+G_{12r}\Gamma[0,m\frac{r}{a}]+H_{12}\cos2\theta+J_{12}+K_{12},
\end{equation}
with incomplete Gamma function $\Gamma[0,z]=E_1(z)$ and with coefficients
\begin{eqnarray}
\label{A.2}G_{120}&=&\frac{im^6e^m}{430080(1+m)ar^3}\bigg[3m^2(-120+7m^2)a^4+3(896+168m^2-9m^4)a^2r^2-8064r^4\nonumber\\
&+&5m^2\bigg((-120+7m^2)a^4+3 (56-3m^2)a^2r^2\bigg)\cos2\theta\bigg],\nonumber\\
G_{12r}&=&\frac{im^6e^mr^2}{215040(1+m)a^6}\bigg[3(896a^4-24m^2a^2r^2+m^4r^4)+5m^2(-24a^2r^2+m^2r^4)\cos2\theta\bigg],\nonumber\\
H_{12}&=&\frac{im}{86016(1+m)a^5r^3}\bigg[(18432a^8+4320ma^7r-288m^2a^6r^2-48m^3a^5r^3+48m^4a^4r^4\nonumber\\
&-&36m^5a^3r^5+44m^6a^2r^6+2m^7ar^7-2m^8r^8)e^{m-m\frac{r}{a}}\nonumber\\
&-&(18432+11376m+2160m^2-120m^3-72m^4+78m^5-106m^6-7m^7+7m^8)a^8\nonumber\\
&+&3m(2352+816m-24m^2-40m^3+38m^4-50m^5-3m^6+3m^7)a^6r^2\bigg],\nonumber\\
J_{12}&=&\frac{-im}{143360(1+m)ar^3}\bigg[(18432+11376m+2160m^2-120m^3-72m^4+78m^5-106m^6\nonumber\\
&-&7m^7+7m^8)a^4+m(9072-656m-824m^2+1016m^3-114m^4+150m^5\nonumber\\
&+&9m^6-9m^7)a^2r^2-2688m(2+2m-m^2+m^3)r^4\bigg],\nonumber\\
K_{12}&=&\frac{im}{71680(1+m)a^5r^3}\bigg[9216a^8+7536ma^7r-1936m^2a^6r^2+872m^3a^5r^3\nonumber\\
&-&872m^4a^4r^4-18m^5a^3r^5+22m^6a^2r^6+m^7ar^7-m^8r^8\bigg]e^{m-m\frac{r}{a}}.
\end{eqnarray}
The element $V_{23}$ is given by
\begin{equation}
\label{A.3}V_{23}=G_{230}\Gamma[0,p]+G_{23r}\Gamma[0,p\frac{r}{a}]+H_{23}\cos3\theta+(J_{23}+K_{23})\cos\theta,
\end{equation}
with coefficients
\begin{eqnarray}
\label{A.4}G_{230}&=&\frac{ip^8e^p}{14192640(1+p)r^4}\bigg[\bigg(p^2(385-9p^2)a^4-11(864+45p^2-p^4)a^2r^2+15840r^4\bigg)\cos\theta\nonumber\\
&+&7p^2\bigg((385-9p^2)a^4-11(45-p^2)a^2r^2\bigg)\cos\theta\cos2\theta\bigg],\nonumber\\
G_{23r}&=&\frac{-ip^8e^pr^3}{7096320(1+p)a^7}\bigg[(3168a^4-55p^2a^2r^2+p^4r^4)\cos\theta-(385p^2a^2r^2-7p^4r^4)\cos\theta\cos2\theta\bigg],\nonumber\\
H_{23}&=&\frac{ip}{4055040(1+p)a^6r^4}\bigg[(-599040a^{10}-171360pa^9r+1440p^2a^8r^3+3120p^3a^7r^3-1200p^4a^6r^4\nonumber\\
&+&420p^5a^5r^5-172p^6a^4r^6+98p^7a^3r^7-106p^8a^2r^8-2p^9ar^9+2p^{10}r^{10})e^{p-p\frac{r}{a}}\nonumber\\
&+&(599040+385200p+85680p^2+840p^3-2760p^4\nonumber\\
&+&1230p^5-554p^6+331p^7-367p^8-9p^9+9p^{10})a^{10}\nonumber\\
&-&11p(19440+7920p+360p^2-360p^3+150p^4-66p^5+39p^6-43p^7-p^8+p^9)a^8r^2\bigg],\nonumber\\
J_{23}&=&\frac{ip}{3153920(1+p)r^4}\bigg[(599040+385200p+85680p^2+840p^3-2760p^4\nonumber\\
&+&1230p^5-554p^6+331p^7-367p^8-9p^9+9p^{10})a^4\nonumber\\
&+&11p(3600-3312p-1512p^2+744p^3-342p^4+258p^5-39p^6+43p^7+p^8-p^9)a^2r^2\nonumber\\
&-&3520p(24+24p-6p^2+2p^3-p^4+p^5)r^4\bigg],\nonumber\\
K_{23}&=&\frac{-ip}{1576960(1+p)a^6r^4}\bigg[299520a^{10}+170160pa^9r-17616p^2a^8r^2+2664p^3a^7r^3\nonumber\\
&-&808p^4a^6r^4+494p^5a^5r^5-618p^6a^4r^6-49p^7a^3r^7+53p^8a^2r^8+p^9ar^9-p^{10}r^{10}\bigg]e^{p-p\frac{r}{a}}.\nonumber\\
\end{eqnarray}
Finally, the element $V_{22}$ is given by
\begin{eqnarray}
\label{A.5}V_{22}&=&\Re\bigg[G^+_{220}\Gamma[0,p]+G^+_{22r}\Gamma[0,p\frac{r}{a}]+G^-_{220}\Gamma[0,m]+G^-_{22r}\Gamma[0,m\frac{r}{a}]\bigg]\nonumber\\
&+&G_{2202}\Gamma[0,2s]+G_{22r2}\Gamma[0,2s\frac{r}{a}]+H_{22}\cos3\theta+J_{22}\cos\theta,\nonumber\\
\end{eqnarray}
with coefficients
\begin{eqnarray}
\label{A.6}G^+_{220}&=&\frac{-im^8e^{im}}{28385280r^4}\frac{6+6m+3m^2+m^3}{1+2s+2s^2}\bigg[(616+9m^2)a^4-11(72+m^2)a^2r^2\bigg](9\cos\theta+7\cos3\theta),\nonumber\\
G^+_{22r}&=&\frac{-im^8e^{im}r^5}{14192640a^7}\frac{6+6m+3m^2+m^3}{1+2s+2s^2}(88a^2+m^2r^2)(9\cos\theta+7\cos3\theta),\nonumber\\
G^-_{220}&=&\frac{ip^6e^m}{14192640r^4}\frac{6+6p+3p^2+p^3}{1+2s+2s^2}\bigg[\bigg(p^2(154+9p^2)a^4+11(1728-18p^2-p^4)a^2r^2\nonumber\\
&-&31680r^4\bigg)\cos\theta+7p^2\bigg((154+9p^2) a^4-11(18+p^2)a^2r^2\bigg)\cos\theta\cos2\theta\bigg],\nonumber\\
G^-_{22r}&=&\frac{ip^6e^mr^3}{7096320a^7}\frac{6+6p+3p^2+p^3}{1+2s+2s^2}\bigg[(6336a^4+22p^2a^2r^2+p^4r^4)\cos\theta\nonumber\\
&+&7p^2(22a^2r^2+p^2r^4)\cos\theta\cos2\theta\bigg],\nonumber\\
G_{2202}&=&\frac{s^6e^{2s}}{1155(1+2s+2s^2)r^4}\bigg[\bigg(9s^4a^4-11(27+s^4)a^2r^2+495r^4\bigg)\cos\theta\nonumber\\
&+&7s^4(9a^4-11a^2r^2)\cos\theta\cos2\theta\bigg],\nonumber\\
G_{22r2}&=&\frac{2s^6e^{2s}r^3}{1155(1+2s+2s^2)a^7}\bigg[(-99a^4+s^4r^4)\cos\theta+7s^4r^4\cos\theta\cos2\theta\bigg],\nonumber\\
H_{22}&=&\frac{1}{63360a^6r^4s(+2s+2s^2}\bigg[(-(88560+177120s+177120s^2+109170s^3+50580s^4\nonumber\\
&+&9540s^5+1098s^6-1074s^7+486s^8-1256s^9+92s^{10}-27s^{11}+18s^{12}+18s^{13})a^{10}\nonumber\\
&+&11s^2(4320+6210s+3780s^2+900s^3+162s^4\nonumber\\
&-&126s^5+54s^6-144s^7+6s^8-3s^9+2s^{10}+2s^{11})a^8r^2\nonumber\\
&+&\bigg((1+s)^2(56160a^{10}-450s^4a^6r^4+159s^8a^2r^8)\nonumber\\
&+&(3+3s+s^2)(10710s^2a^9r+105s^6a^5r^5+2s^{10}ar^9)\nonumber\\
&+&(3+3s-s^3)(390s^3a^7r^3-49s^7a^3s^7)\nonumber\\
&+&(3-3s^2-2s^3)(90s^2a^8r^2+43s^6a^4r^6+2s^{10}r^{10})\bigg)e^{s-s\frac{r}{a}}\cos(s-s\frac{r}{a})\nonumber\\
&+&\bigg((1+s)^2(270s^2a^8r^2+129s^6a^4r^6+6s^{10}r^{10})\nonumber\\
&+&(3+3s+s^2)(390s^4a^7r^3-49^8a^3r^7)-(3+3s-s^3)(10710sa^9r+105s^5a^5r^5+2s^9ar^9)\nonumber\\
&-&(3-3s^2-2s^3)(18720a^{10}-150s^4a^6r^4+53s^8a^2r^8)\bigg)e^{s-s\frac{r}{a}}\sin(s-s\frac{r}{a})\nonumber\\
&+&24(1350a^{10}+2700sa^9r+1710s^2a^8r^2+315s^3a^7r^3-90s^4a^6r^4\nonumber\\
&+&30s^5a^5r^5-12s^4a^6r^4+6s^7a^3r^7-4s^4a^2r^8+4s^9ar^9-8s^{10}r^{10})e^{2s-2s\frac{r}{a}}\bigg],\nonumber\\
\end{eqnarray}
\begin{eqnarray}
J_{22}&=&\frac{1}{49280a^6r^4s(1+2s+2s^2)}\bigg[880s^3(18+24s+12s^2-12s^3+3s^4-2s^5-2s^6)a^6r^4\nonumber\\
&+&11s^2(1728+2466s+1476s^2+324s^3+738s^4-270s^5\nonumber\\
&+&150s^6-48s^7+6s^8-3s^9+2s^{10}+2s^{11})a^8r^2\nonumber\\
&-&(88560+177120s+177120s^2+109170s^3+50580s^4+9540s^5+1098s^6-1074s^7\nonumber\\
&+&486s^8-1256s^9+94s^{10}-27s^{11}+18s^{12}+18s^{13})a^{10}\nonumber\\
&+&\bigg((1+s)^2(56160a^{10}+606s^4a^6r^4+159s^8a^2r^8)\nonumber\\
&+&(3+3s+s^2)(21270s^2a^9r-247s^6a^5r^5+2s^{10}ar^9)\nonumber\\
&-&(3+3s-s^3)(666s^3a^7r^3+49s^7a^3r^7)\nonumber\\
&+&(3-3s^2-2s^3)(2202s^2a^8r^2-309s^6a^4r^6+2s^{10}r^{10})\bigg)e^{s-s\frac{r}{a}}\cos(s-s\frac{r}{a})\nonumber\\
&+&\bigg((1+s)^2(6606s^2a^8r^2-927s^6a^4r^6+6s^{10}r^{10})-(3+3+s^2)(666s^4a^7r^3+49s^8a^3r^7)\nonumber\\
&-&(3-3s^2-2s^3)(18720a^{10}+202s^4a^6r^4+53s^8a^2r^8)\nonumber\\
&-&(3+3s-s^3)(21270sa^9r-247s^5a^5r^5+2s^9ar^9)\bigg)e^{s-s\frac{r}{a}}\sin(s-s\frac{r}{a})\nonumber\\
&+&24\bigg(1350a^{10}+2700sa^9r+1314s^2a^8r^2+183s^3a^7r^3\nonumber\\
&-&2s^4a^6r^4-58s^5a^5r^5+164s^6a^4r^6+6s^7a^3r^7-4s^8a^2r^8+4s^9ar^9-8s^{10}r^{10}\bigg)e^{2s-2s\frac{r}{a}}\bigg].\nonumber\\
\end{eqnarray}

The complexity of the above expressions reflects that of the Reynolds force density in Eq. (6.4) compounded with that of the Green function in Eq. (7.6).

\newpage

\newpage

\section*{Figure captions}

\subsection*{Fig. 1}
Plot of the reduced swimming velocity $U_{red}(s)$ as a function of scale number $s$ for the $B_1B_2$-swimmer with mode coefficients listed in Eq. (4.2) with $B_2=5B_1$ (long dashes), compared with that for $12$-swimmer with optimized coefficients $(\mu^I_1,\kappa^I_2,\mu^I_2)=(1,-4i\sqrt{2}/3,11i/(5\sqrt{2}))$ (solid curve). We also plot the reduced swimming velocity $U_{red}(s)$ as a function of scale number $s$ for the $123$-swimmer with mode coefficients listed in Eq. (4.5) (short dashes).

\subsection*{Fig. 2}
Plot of the reduced swimming velocity $U_{red}(s)$ as a function of scale number $s$ for the $12$-swimmer with mode coefficients $\mu^I_1=1,\kappa^I_2=i\sqrt{10}/3,\mu^I_2=0$ (solid curve), for the $123$-swimmer with mode coefficients listed in Eq. (4.3) (long dashes), and for the $23$-swimmer with mode coefficients listed in Eq. (4.4) (short dashes).

\subsection*{Fig. 3}
Net flow pattern for potential $12$-swimmer with stroke specified by coefficients $\mu^I_1=1,\kappa^I_2=0,\mu^I_2=i/\sqrt{2}$. The flow is independent of scale number $s$.

\subsection*{Fig. 4}
Net flow pattern for $12$-swimmer with stroke specified by coefficients $\mu^I_1=1,\kappa^I_2=-4i\sqrt{2}/3,\mu^I_2=11i/(5\sqrt{2})$ for scale number $s=0.1$.

\subsection*{Fig. 5}
Same as in Fig. 4 for scale number $s=10$.

\subsection*{Fig. 6}
Plot of the $z$-component of the net flow velocity in the equatorial plane $\theta=\pi/2$ for the swimmer of Fig. 5, separately for $v'_{Sz}$ (solid curve) and $v'_{Vz}$ (dashed curve).

\subsection*{Fig. 7}
Net flow pattern for $123$-swimmer with stroke specified by the coefficients in Eq. (4.5) for scale number $s=10$.

\newpage
\setlength{\unitlength}{1cm}
\begin{figure}
 \includegraphics{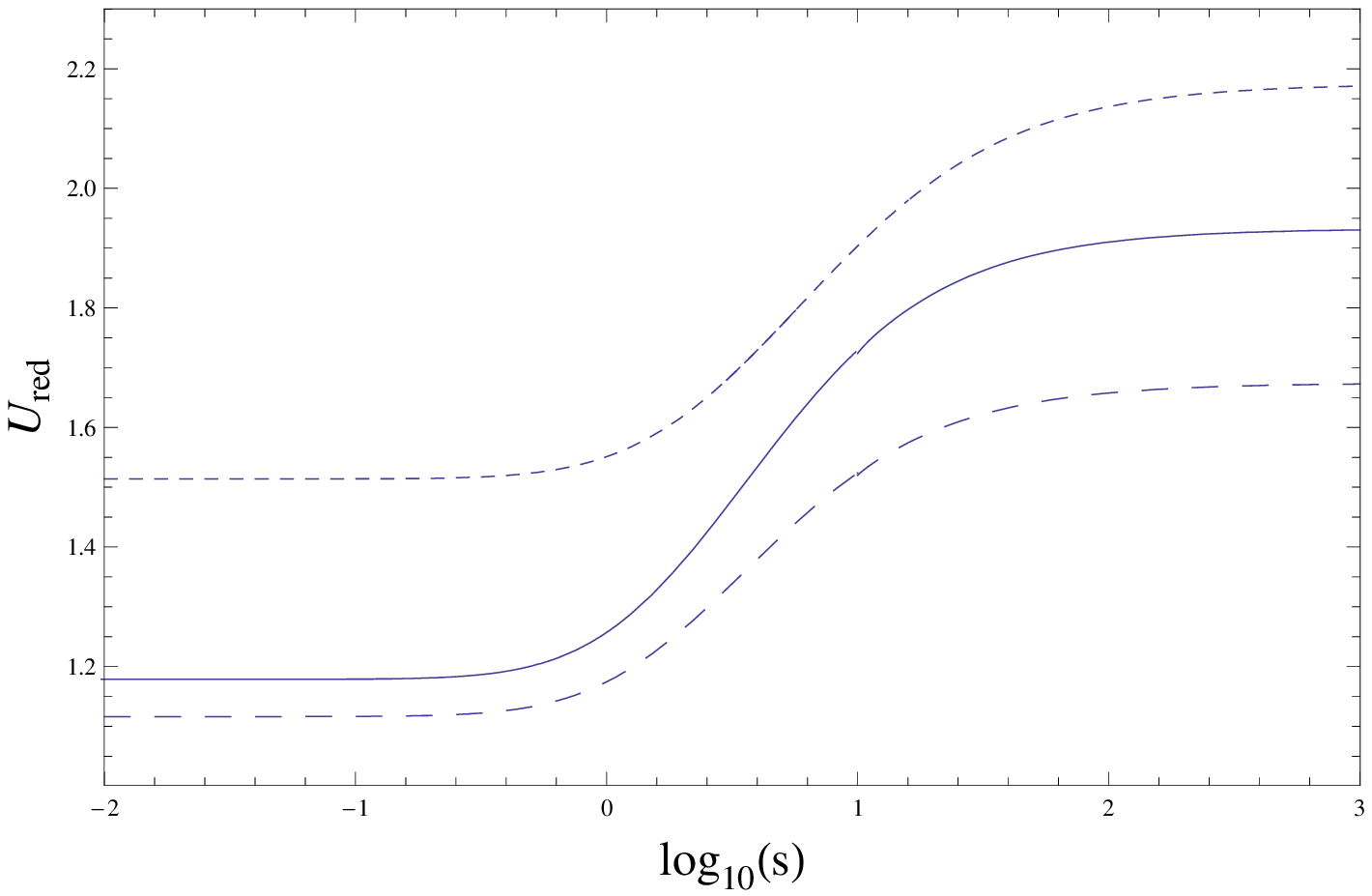}
   \put(-9.1,3.1){}
\put(-1.2,-.2){}
  \caption{}
\end{figure}
\newpage
\clearpage
\newpage
\setlength{\unitlength}{1cm}
\begin{figure}
 \includegraphics{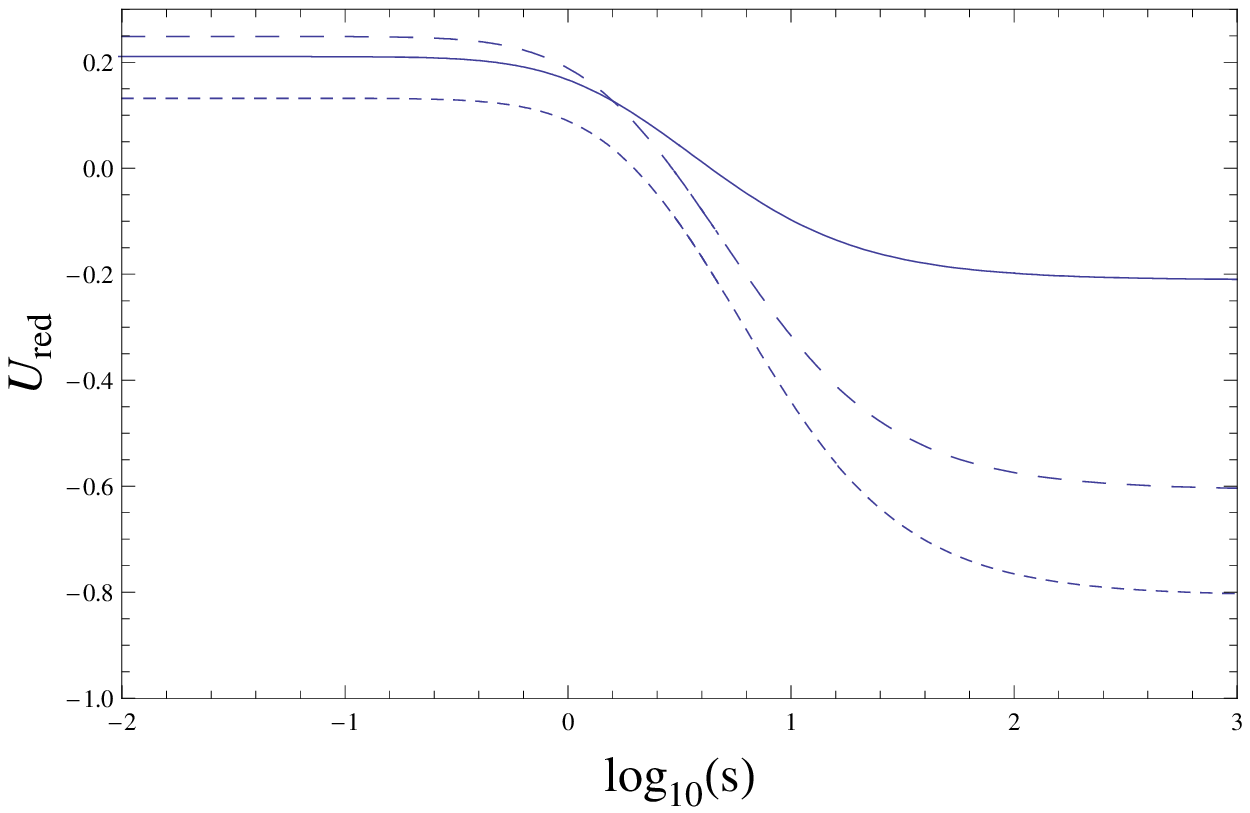}
   \put(-9.1,3.1){}
\put(-1.2,-.2){}
  \caption{}
\end{figure}
\newpage
\clearpage
\newpage
\setlength{\unitlength}{1cm}
\begin{figure}
 \includegraphics{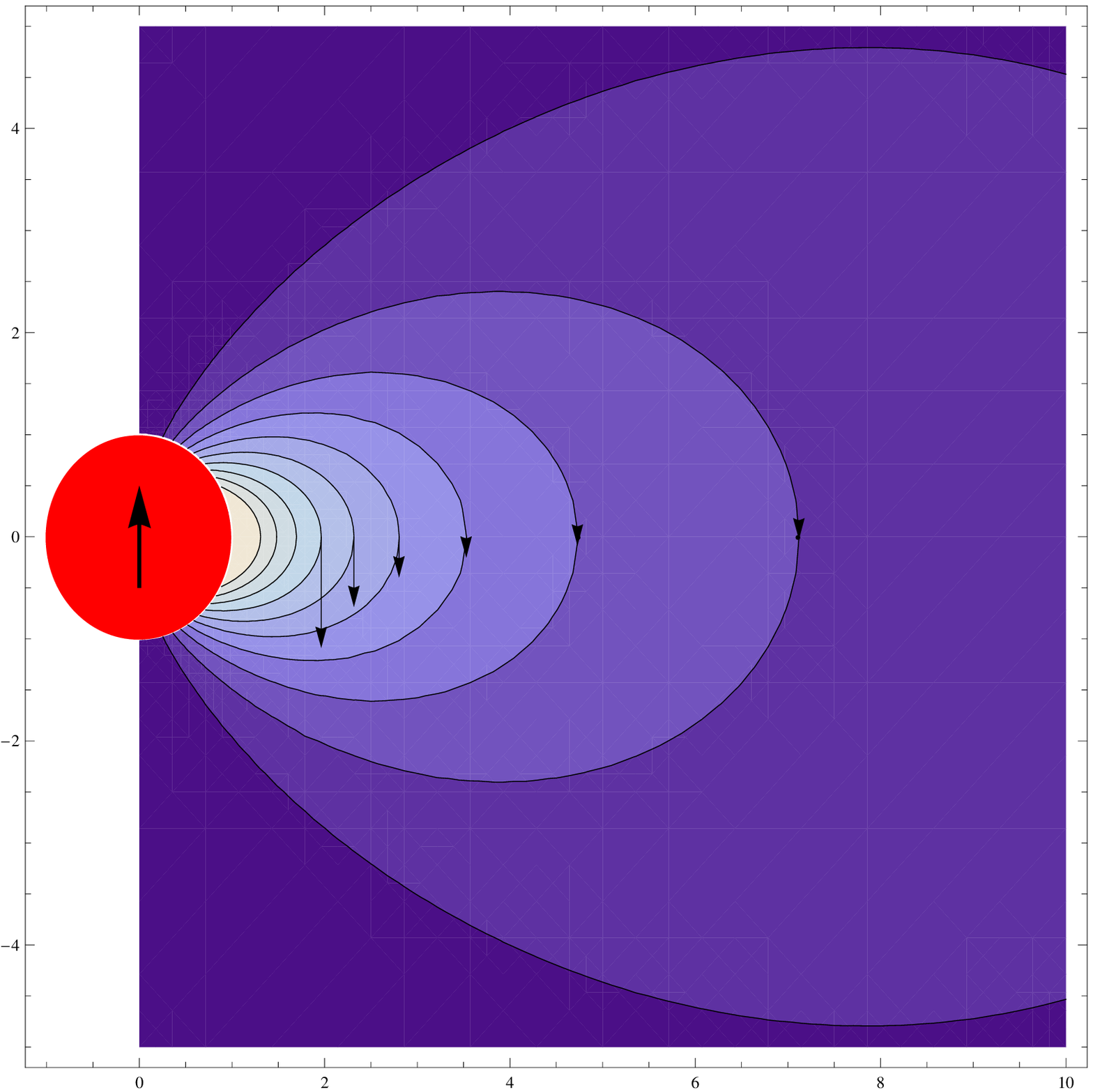}
   \put(-9.1,3.1){}
\put(-1.2,-.2){}
  \caption{}
\end{figure}
\newpage
\clearpage
\newpage
\setlength{\unitlength}{1cm}
\begin{figure}
 \includegraphics{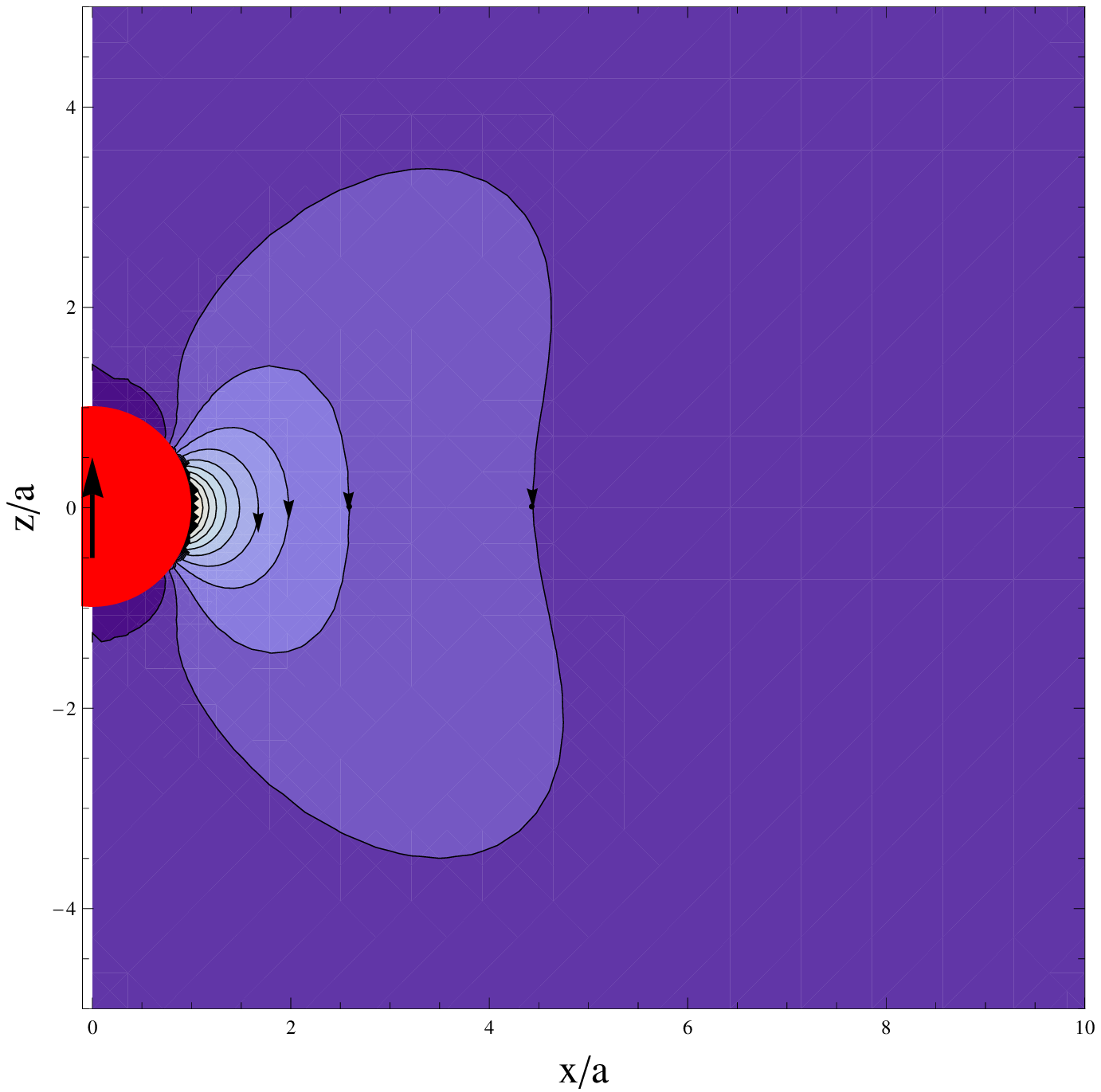}
   \put(-9.1,3.1){}
\put(-1.2,-.2){}
  \caption{}
\end{figure}
\newpage
\clearpage
\newpage
\newpage
\setlength{\unitlength}{1cm}
\begin{figure}
 \includegraphics{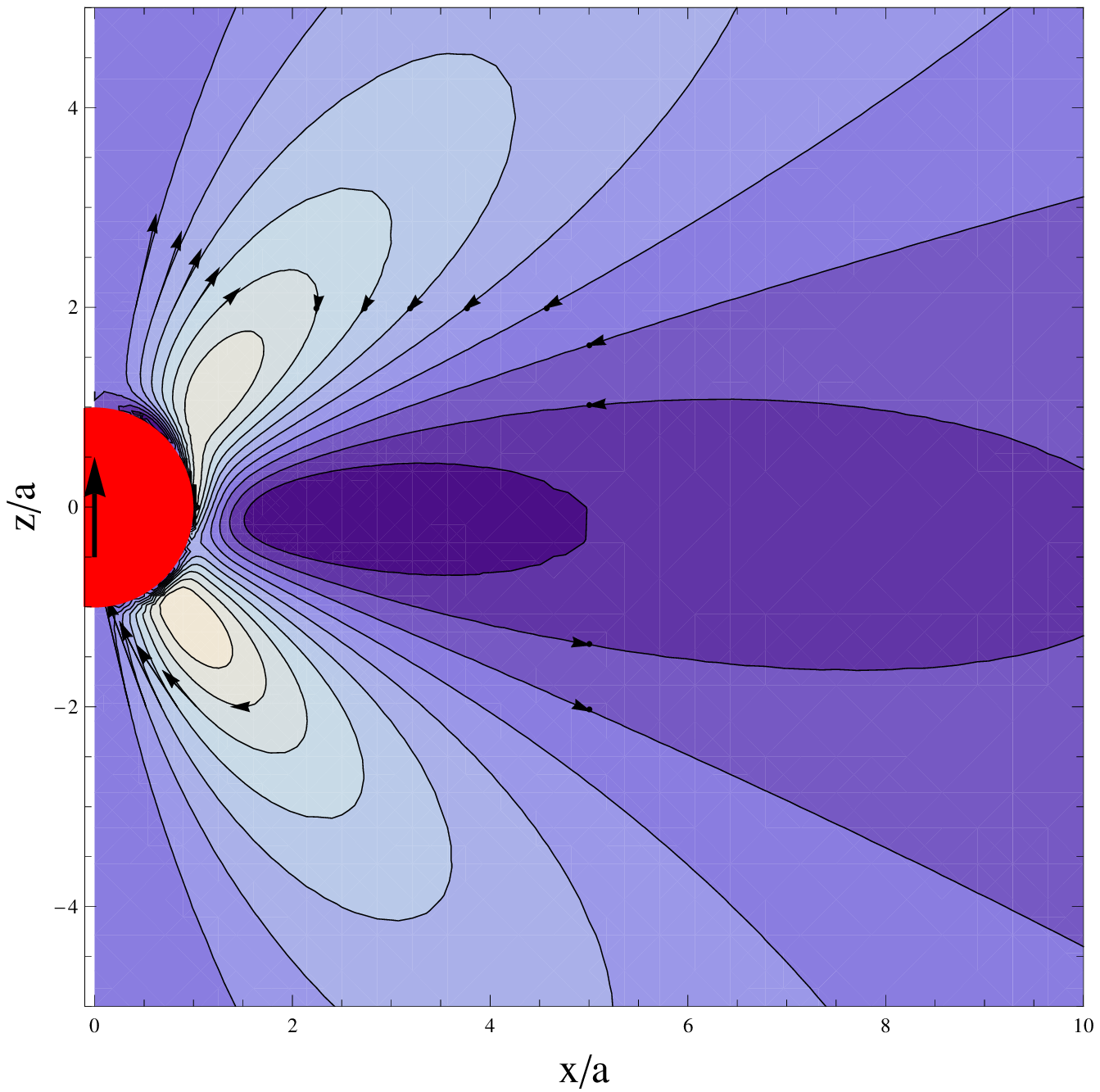}
   \put(-9.1,3.1){}
\put(-1.2,-.2){}
  \caption{}
\end{figure}
\newpage
\clearpage
\newpage
\setlength{\unitlength}{1cm}
\begin{figure}
 \includegraphics{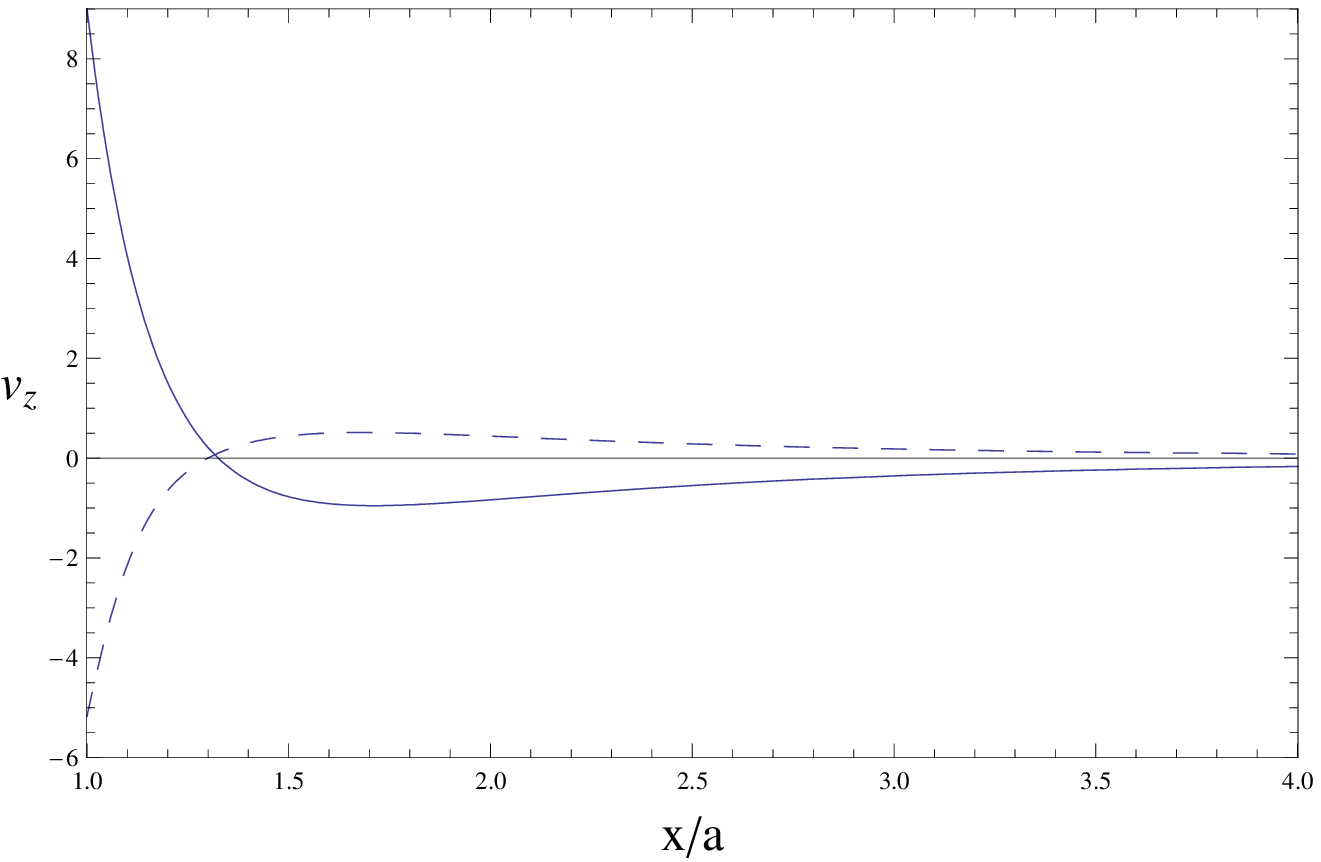}
   \put(-9.1,3.1){}
\put(-1.2,-.2){}
  \caption{}
\end{figure}
\newpage
\clearpage
\newpage
\newpage
\setlength{\unitlength}{1cm}
\begin{figure}
 \includegraphics{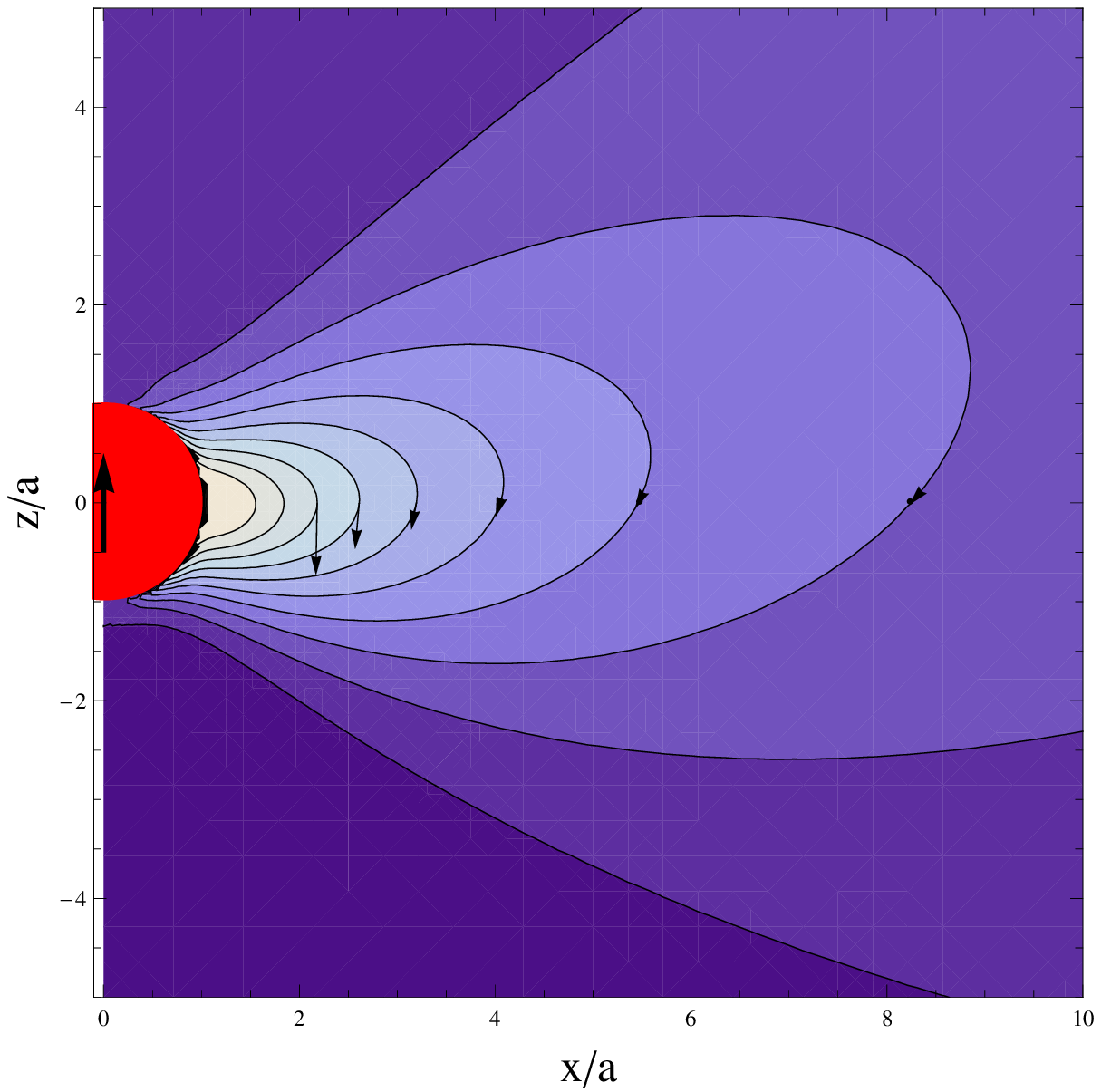}
   \put(-9.1,3.1){}
\put(-1.2,-.2){}
  \caption{}
\end{figure}
\newpage
\clearpage
\newpage
\end{document}